\documentclass[prd,tightenlines,nofootinbib,superscriptaddress]{revtex4}

\usepackage[T1]{fontenc}
\usepackage[utf8]{inputenc}

\usepackage{amsfonts,amssymb,amsthm,bbm,amsmath,wasysym}
\usepackage{hyperref}

\usepackage{subcaption}
\usepackage{color,psfrag}
\usepackage[dvips]{graphicx}

\usepackage{tikz}
\usetikzlibrary{calc}
\usetikzlibrary{decorations.pathmorphing}

\newcommand{\SU}{\mathrm{SU}}
\newcommand{\ISU}{\mathrm{ISU}}

\newcommand{\be}{\begin{equation}}
\newcommand{\ee}{\end{equation}}
\newcommand{\beq}{\begin{eqnarray}}
\newcommand{\eeq}{\end{eqnarray}}
\newcommand{\bes}{\begin{eqnarray}}
\newcommand{\ees}{\end{eqnarray}}

\newcommand{\id}{{\mathbb{I}}}

\begin{document}

\title{Simplicity constraints: a 3d toy-model for Loop Quantum Gravity}

\author{{\bf Christoph Charles}}\email{christoph.charles@univ-lyon1.fr}
\affiliation{Univ Lyon, Université Lyon 1, CNRS/IN2P3, IPN-Lyon, F-69622, Villeurbanne, France}

\date{\today}

\begin{abstract}
  In Loop Quantum Gravity, tremendous progress has been made using the Ashtekar-Barbero variables. These variables, defined in a gauge-fixing of the theory, correspond to a parametrization of the solutions of the so-called simplicity constraints. Their geometrical interpretation is however unsatisfactory as they do not constitute a space-time connection. It would be possible to resolve this point by using a full Lorentz connection or, equivalently, by using the self-dual Ashtekar variables. This leads however to simplicity constraints or reality conditions which are notoriously difficult to implement in the quantum theory.
  
  We explore in this paper the possibility of imposing such constraints at the quantum level in the context of canonical quantization. To do so, we define a simpler model, in 3d, with similar constraints by extending the phase space to include an independent vielbein. We define the classical model and show that a precise quantum theory by gauge-unfixing can be defined out of it, completely equivalent to the standard 3d euclidean quantum gravity.
  
  We discuss possible future explorations around this model as it could help as a stepping stone to define full-fledged covariant Loop Quantum Gravity.
\end{abstract}

\maketitle

\tableofcontents

\section{Introduction}

Loop Quantum Gravity (LQG) is based on a reformulation of Einstein theory of general relativity as a gauge theory over an $\SU(2)$ connection called the Ashtekar-Barbero connection \cite{Ashtekar:1986yd, Barbero:1994ap, Immirzi:1996di}. The symmetries of the theory naturally splits into the $\SU(2)$ gauge symmetry and the diffeomorphism symmetry of general relativity \cite{Ashtekar:1987gu}. This means that Loop Quantum Gravity does not have explicit Lorentz symmetry. And indeed, the natural formulation of LQG can be derived from a gauge-fixing (called the time-gauge) of first-order general relativity with a Holst term \cite{Holst:1995pc}. The $\SU(2)$ gauge symmetry therefore corresponds to the rotational part of the local Lorentz symmetry. One of the profound consequences of this choice of variable is that the $\SU(2)$ connection is a spatial connection and not a space-time one \cite{Samuel:2000ue}. This may cast doubt on the correct behaviour of the theory regarding space-time symmetries.

It has been argued that the correct behaviour for the theory should be obtained by using true space-time connections \cite{Samuel:2000ue, Alexandrov:2001wt, Alexandrov:2002br, Achour:2015zmk, Achour:2014rja}. In this regard, the definition of the Ashtekar-Barbero connection depends on a, \textit{a priori} complex, parameter called the Immirzi parameter $\gamma$ \cite{Immirzi:1996di}. In the Lorentzian case, when $\gamma = \pm \mathrm{i}$, the Ashtekar-Barbero connection reduces to the original Ashtekar connection, which is the pullback of a space-time connection and is the self-dual (or anti-self-dual depending on the sign of $\gamma$) part of the Lorentz spin connection of the manifold. Therefore, there is a nice space-time geometrical interpretation of the variables for $\gamma = \pm\mathrm{i}$. The problem with this choice of the variables is that they are complex, which requires the introduction of additional reality conditions. These conditions are difficult to implement even though they have the courtesy of having a nice interpretation as second class constraints enforcing the so-called simplicity conditions \cite{Alexandrov:2005ng}.

\medskip

Therefore, a different strategy might be to try and solve these constraints, for instance using the Dirac algorithm. The new Dirac brackets could then be quantized as a commutator and the second class constraints be implemented as operator constraints. There have been very interesting results in this direction \cite{Alexandrov:2001wt}. It turns out that two sets of variables are preferred with respect to the Dirac brackets. The first set is a space-time Lorentz connection (and its conjugate) which is however non-commutative with respect to the brackets. The second set is a spatial Lorentz connection (and its momentum), which is not a pull-back of a space-time connection, but which is commutative with respects to the brackets. This last connection naturally generalizes the Ashtekar-Barbero connection to the Lorentzian symmetry and reduces to it in the time-gauge. It can even be shown that a nice parametrization of the constraint manifold comes from the usual Ashtekar-Barbero connection completed with a 3d vector (and all the conjugates) which mix into a full spatial Lorentz connection \cite{BarroseSa:2000vx}. This parametrization has a very nice Hamiltonian structure though the space-time interpretation is less appealing.

This means that from this perspective two sets of variables are natural to describe quantum gravity and that the choice between them will boil down to a trade-off: either we have commutativity of the variables or we have a nice space-time geometrical interpretation of them. Choosing the geometrical interpretation may seem more natural but leads to a lot of technical difficulties linked to the non-commutativity of the brackets. But choosing the commutativity, though technically simpler, might lead to an anomalous theory with regard to space-time diffeomorphisms. Some explorations support however the point that the resulting theory does behave nicely. For instance, this generalized connection can be used in a space-gauge (similar to a time-gauge but with 2+1 Minkowski space-time sheets) and lead to the same spectrum for the usual observables of LQG \cite{Liu:2017bfk}. Recent work \cite{Wieland:2017cmf} also shows that the results from the standard quantization, in particular the discrete spectrum of the area, can be recovered on the boundary without the time-gauge or an unnatural choice of variable. This gives support to the idea that the standard quantization is not anomalous.

\medskip

The choice of variables is also linked to the problem of the role of the Immirzi parameter. Indeed, the Hamiltonian structure for the two sets of variables discovered by Alexandrov \cite{Alexandrov:2001wt} is not the same. The Dirac brackets for the non-commutative connection do not involve the Immirzi parameter at all whereas the brackets of the commutative variables do. The problem has been investigated by Dittrich \textit{et al.} \cite{Dittrich:2012rj} in a discrete setting. It was shown there that a reduction of the secondary simplicity constraints might be done in two different ways leading to two different Hamiltonian structures. There is a natural match between these two discrete reductions and the choice of continuum variables. This is important because it enlightens the geometric meaning of the different possible variables, and the possible implications for the quantum theories. In particular, the non-commutative variables seem to correspond to the reduction to a Regge geometry whereas the commutative ones only reduce to twisted geometry. This suggests that by using the Ashtekar-Barbero variables, we are actually missing some conditions in the theory, that might not be visible in the classical theory, but does leave its mark in the quantum theory, for instance in the choice of representation. This suggests that the full covariant theory should have a representation for which the natural geometrical interpretation is one of Regge geometries. There is however another possibility which has been explored in \cite{Anza:2014tea} that the reduction can appear through the dynamics or through the preservation of the primary constraints.

\medskip

Ultimately, these questions would be solved by testing different quantization methods and different choices of variables. Because of all the aforementioned difficulties, it is a daunting task in the full theory. Our goal in this paper is therefore to develop a simplified model, with second class constraints, which should mimic the problem from the full theory but in a more tractable context. In practice, we will develop techniques in the solvable model of 3d discrete quantum gravity. In this paper, we will define an extension of the usual 3d quantum gravity model that incorporates second class constraints and is still equivalent classically to it. We will also discuss possible quantization methods, as a proof of concept to explore the possible equivalence at the quantum level. As the simplicity constraints in 3d are trivial, we do not expect any difficulty. And indeed, we will reproduce the gauge unfixing method already tested in the full theory \cite{Bodendorfer:2011nv} to show that the theory can make sense in the quantum setting. Having introduced this model, we will discuss how it could be used to explore various quantization techniques, in particular how it should be useful as a stepping stone for implementing reality conditions in a diffeomorphism invariant theory and develop a possible Regge-like representation of Loop Quantum Gravity.

The main idea of our paper is to consider the usual action of first order 3d gravity as a completely singular action. This means that we consider the usual variables, that is the vielbein $e$ and the spin connection $\omega$ as independent variables with constraints on (all) their corresponding momenta. This is actually quite straightforward and what the usual Dirac procedure leads to. The resulting Hamiltonian theory has naturally new constraints to compensate for the additional variables. These constraints are second class and it appears that the standard approach really correspond to the classical solving of these constraints. The opportunity for us, however, is that the unsolved system is a relatively simple diffeomorphism invariant theory that still carries second class constraints. It is therefore a good starting point to explore quantization techniques for second class system in a diffeomorphism invariant setting. We explore one possible quantization, but argue that the model should be used as a testbed for others.

The construction follows some developments in \cite{Randono:2008bb} by enlarging the phase space with independent triad variables. Similar techniques have recently been developed in \cite{Belov:2017who}. It should be noted however that the focus is way different. In \cite{Randono:2008bb}, the goal is to have a natural parametrization of the phase space that removes the conjugated momenta and reveals a new gauge group. Our approach may be more down-to-earth as we just want to develop a simple diffeomorphism invariant model with second class constraints. Similarly the focus of \cite{Belov:2017who} is more spinfoam techniques whereas our paper is fully done in the canonical setting. Still, the papers should be complementary as we insist on the same point: enlarging the phase space is natural when dealing with the simplicity constraints.

Though the goal is similar to \cite{Geiller:2011aa} as we are trying to elucidate the simplicity constraints using 3d quantum gravity as a testing environment, the gist of the paper is quite different. In their paper, Geiller and Noui have introduced a theory that reduces to 3d quantum gravity via non-trivial simplicity constraints. In our paper, we rather try to understand the simplicity constraints by studying full-fledged, but vanilla, 3d quantum gravity, hoping that the insistance on keeping the triad variables might help. We do believe that these approaches are complementary and that both can help studying the full 4d theory.

We will now have a bird eye view over the technique we want to implement. The rest of the article will develop the more technical aspects of it. Our goal is to implement the simplicity constraints that naturally appear in general relativity at the quantum level, following canonical quantization techniques. The usual constraints are quite difficult to implement as they involve highly non-linear (not even polynomial) expressions. Even in the simplest case, that is 4d euclidean gravity, with the time gauge, the constraints read:
\begin{equation}
\left\{\begin{array}{rcl}
E_i^{(+)a} &=& E_i^{(-)a}\,, \\
A_a^{(+)i} - \Gamma_a^i(E^{(+)}) &=& -A_a^{(-)i} + \Gamma_a^i(E^{(-)})\,.
\end{array}\right.
\end{equation}
Here, the $A^{(\pm)}$ are the self-dual and anti-self-dual part of the Ashtekar connection.
The $E^{(\pm)}$ are the conjugated momenta associated to $A^{(\pm)}$. They will, thanks to the first constraint, fall back to the single densitized triad. Finally, $\Gamma(E)$ is the unique spin connection without torsion compatible with the metric and the densitized triad $E$. The second constraint imply that the $A^{(\pm)}$ are constructed from the spin connection associated to the densitized triad and with opposite contribution from a quantity which turns out to be the extrinsic curvature $K_a^i$. In equations, this reads: $A_a^{(\pm)i} = \Gamma_a^i(E^{(\pm)}) \pm K^i_a$ . This second constraint is the source of problem. The appearance of $\Gamma$ makes it not polynomial. The quantization is therefore extremely hard to do.

A restatement of the idea of this paper is to implement these secondary constraints by extending the phase space of general relativity. Instead of searching for conditions that imply the existence of some field $e^I$, we will extend the phase space of general relativity to include such fields and impose conditions so that $B$ can be expressed with them. In the full 3+1d Lorentzian theory, such a splitting has been studied \cite{Randono:2008bb} and might be helpful. But our interest is that it generalizes to every dimension and in particular to three dimensions. As we mentioned, such an approach is not far-fetched as it is natural when we consider the usual first-order action of gravity. When keeping $e$ (the vielbein) and $\omega$ (the connection) as independent variables, the condition $B^{IJ} = e^I \wedge e^J$ (plus a possible Holst term) will appear as a primary constraint in the Hamiltonian analysis. Then, the full Hamiltonian analysis will reveal all the constraints we need to impose.

Let's therefore explore the idea in the simplified context of 3d gravity. At first, this might be a little silly as there are no simplicity constraints in 3d. We want to highlight however that it is still possible to consider $\omega$ and $e$ as the independent fundamental degrees of freedom. This will lead to the trivial simplicity constraint $B = \star e$. Still, eventhough the condition is trivial, it will be part of a second class constraint system on which we can test our techniques. Another advantage of 3d quantum gravity comes from its topological nature. Thanks to this absence of local degrees of freedom, understanding the theory on a fixed graph is sufficient to understand the theory in the continuum. This comes in handy, as imposing the simplicity constraints only on a fixed graph solves all the difficulties linked to diffeomorphism invariance. This will allow to test the method with various methods and check how they might be generalized to a continuous setting.

We will develop the corresponding technical points in the rest of this paper. It is organized in the following manner. The paper is divided into three sections. In this first section, we recall the standard analysis and then develop our own. We also discuss the problem of finding a discrete subsystem for quantization. In the second section, we concentrate on the corresponding quantum theory. We discuss the problem of imposing the second class constraints and explore one possible avenue. We then define the Hilbert space, the observables to finally define the quantum theory and check that it is equivalent to the standard analysis. In a final section, we discuss possible future explorations. We consider two ideas more precisely: we discuss how our Hamiltonian analysis might be adapted to the 3+1d case and how this might help develop a Regge-like representation for Loop Quantum Gravity. The second idea we explore is possible modification of the model to get closer to the problem of reality conditions similar to the one we have for self-dual variables. The extra forth section concludes the paper. We also added an appendix detailing the full Hamiltonian analysis in 3+1d but going no further (in particular we do not gauge-unfix or pursue the computation of Dirac brackets).

\section{Classical setup}

\subsection{Standard approach}

As we hinted in the introduction, our goal is find a problem analogous to the simplicity constraints in 3d to use it as a toy-model before heading off to the full problem. But before dwelling into the details, let's recall how the usual analysis goes for 3d quantum gravity. We will therefore sketch the traditional Hamiltonian analysis, the choice of discrete variables and have some brief comments on the quantization of such a structure.

As usual for 3d quantum gravity, we start from the following action (using Planck units):
\begin{equation}
S[e,\omega] = \frac{1}{2} \int_\mathcal{M} \epsilon_{IJK} e^I \wedge F^{JK}[\omega].
\end{equation}
where $e$ is the triad and $\omega$ is a $\mathrm{SU}(2)$ connection. Note that we used the identification between $\mathfrak{su}(2)$ and the 2-contravariant tensors in $3$ dimensions. The integration is done over a differentiable manifold $\mathcal{M}$ which we assume to be orientable. We will assume it to be isomorphic to $\mathbb{R}\times \Sigma$ where $\Sigma$ is a 2-dimensional manifold with no boundary (\textit{i.e.} $\partial \Sigma = \varnothing$). $F$ represents the curvature of $\omega$ and reads:
\begin{equation}
F^{IJ}_{ab} = \partial_a \omega^{IJ}_{b} - \partial_b \omega^{IJ}_{a} + \omega^{IK}_{a} \omega^{LJ}_{b} \eta_{KL} - \omega^{IK}_{b} \omega^{LJ}_{a}\eta_{KL},
\end{equation}
where we have made the manifold indices explicit. Note that the keeping of covariant and contravariant indices is not strictly necessary (as we are dealing with euclidean signature) but we keep it for clarity. Therefore, in a more explicit manner, the action reads:
\begin{equation}
S[e,\omega] = \frac{1}{2} \int_\mathbb{R} dt \int_\Sigma d^2 \sigma \left(\epsilon^{abc} \epsilon_{IJK} e^I_a \left[\partial_b \omega^{JK}_{c} - \partial_c \omega^{JK}_{b} + \omega^{JM}_{b} \omega^{NK}_{c} \eta_{MN} - \omega^{JM}_{c} \omega^{NK}_{b}\eta_{MN}\right]\right).
\end{equation}

Before going to the Hamiltonian theory, it is helpful to rewrite the Lagrangian by separating clearly the time index from the spatial ones. This leads to:
\begin{eqnarray}
S[e,\omega] &=& \frac{1}{2} \int_\mathbb{R} dt \int_\Sigma d^2 \sigma \left(
2\epsilon^{ab} \epsilon_{IJK} e^I_b \partial_0 \omega^{JK}_{a} +
2\epsilon^{ab} \epsilon_{IJK} \partial_a e^I_b  \omega^{JK}_{0}
+ 2\epsilon^{ab} \epsilon_{IJK} e^I_a \omega^{JM}_{b} \omega^{NK}_{0} \eta_{MN} \right. \\
&+& \left.\epsilon^{ab} \epsilon_{IJK} e^I_0 \left[\partial_a \omega^{JK}_{b} - \partial_b \omega^{JK}_{a}  \omega^{JM}_{b} \omega^{NK}_{c} \eta_{MN} - \omega^{JM}_{c} \omega^{NK}_{b}\eta_{MN}\right]\right), \nonumber
\end{eqnarray}
which can be written in a more compact way as follows:
\begin{equation}
S[e,\omega] = \int_\mathbb{R} dt \int_\Sigma d^2 \sigma \left(
\epsilon^{ab} \epsilon_{IJK} e^I_b \partial_0 \omega^{JK}_{a} +
\epsilon^{ab} \epsilon_{IJK} \omega^{JK}_{0} \nabla_a e^I_b +
\frac{1}{2}\epsilon^{ab} \epsilon_{IJK} e^I_0 F^{JK}_{ab}\right).
\end{equation}
where $\nabla$ denotes the covariant derivative.

We can now start developing the Hamiltonian theory as it can be practically read of the previous equation. $\omega_0^{IJ}$ and $e_0^I$ act as Lagrange multipliers, while the Poisson structure is given by:
\begin{equation}
\{\omega^{JK}_a(x),e^I_b(y)\} = \epsilon_{ab}\epsilon^{IJK}\delta^{(2)}(x-y).
\end{equation}
The theory is therefore totally constrained and the constraints read:
\begin{equation}
\nabla_a e^I_b = 0,\quad F^{IJ}_{ab} = 0.
\end{equation}
The first set of constraints constitute the Gauß constraints. The second set implies flatness of the manifold and can be projected, when the triad is non-degenerate, onto the diffeomorphism and Hamiltonian constraints.

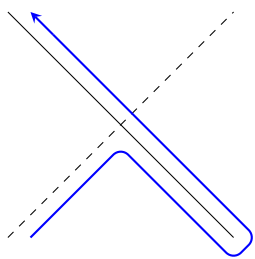
\begin{figure}[h!]
	
	\centering
	
	\begin{tikzpicture}[scale=1.5]
	\coordinate (A) at (0,0);
	\coordinate (B) at (2,0);
	\coordinate (C) at (0,-2);
	\coordinate (D) at (2,-2);
	
	\coordinate (C1) at (0.2,-2);
	\coordinate (O1) at (1.0,-1.2);
	\coordinate (D1) at (2,-2.2);
	\coordinate (D2) at (2.2,-2);
	\coordinate (A1) at (0.4,-0.2);
	\coordinate (A2) at (0.2,0);
	
	\draw (D) -- (A);
	\draw[dashed] (C) -- (B);
	
	\draw[blue,rounded corners,thick] (C1) -- (O1) -- (D1) -- (D2) -- (A1);
	\draw[blue,thick,->,>=stealth] (A1) -- (A2);
	
	\end{tikzpicture}
	
	\caption{The momenta are naturally carried by the dual graph, with edges represented here as solid lines. The natural integration is along the lines of the original graph (dashed lines). For dual quantities, the transport path must be extended to satisfy nice covariance properties. The line in blue represents the parallel transport needed as we integrate along the solid line.}
	\label{fig:momentumintegration}
	
\end{figure}

Quantizing a diffeomorphism-invariant continuous theory usually involves projective limit techniques. In the particular case of 3d quantum gravity though, this is not needed as the theory is topological. A discrete subsystem is enough to describe all the degrees of freedom of the the continuum theory. In order to prepare our work on our toy model, let's recap how the discretization is done in the standard approach. The connection has a very natural discrete equivalent, the holonomy, which is its path- ordered exponential. The holonomy along an oriented path $\gamma$ is defined by:
\begin{equation}
H_{\gamma} = \tilde{H}(1),\quad \tilde{H}(0) = \id,\quad \frac{\mathrm{d}\tilde{H}}{\mathrm{d}s}(s) = \mathrm{i} \tilde{H}(s) \omega^{IJ}_a(\gamma(s)) J_{IJ} \dot{\gamma}^a .
\end{equation}
where $J_{IJ}$ are the generators of the $\mathrm{SU}(2)$ group (for the Euclidean signature). Similarly we can define an integrated triad along a path. It is quite simply defined by:
\begin{equation}
\mathcal{E}^I_{\gamma} = \int_\gamma (H_{\gamma_x} \triangleright e_a)^I \mathrm{d}x^a ,
\end{equation}
where $\gamma_x$ is the partial path ending at the point $x$ of integration. Note that the parallel transport is necessary to get a covariant quantity out of the integration. It is standard to further parallel transport the triad to the same starting point as the holonomy (see figure \ref{fig:momentumintegration}). This is interesting to get covariant brackets. Apart from this, it does not change the bracket among the triads themselves.

\begin{figure}[h!]
	
	\centering
	
	\begin{tikzpicture}
	
	\def \step {60}
	\def \r {3.0}
	\def \sqrtthree {1.732050808}

	\coordinate (C3) at (0,0);
	\coordinate (C1) at ($(C3) + (0*\step:\r)$);
	\coordinate (C2) at ($(C3) + (1*\step:\r)$);
	\coordinate (C4) at ($(C3) + (2*\step:\r)$);
	\coordinate (C5) at ($(C3) + (3*\step:\r)$);
	
	\coordinate (D1) at ($(C3) + (0.5*\step:\sqrtthree*\r/3.0)$);
	\coordinate (D2) at ($(C3) + (1.5*\step:\sqrtthree*\r/3.0)$);
	\coordinate (D3) at ($(C3) + (2.5*\step:\sqrtthree*\r/3.0)$);

	\draw (C1) -- (C2) -- (C3) -- (C4) -- (C5);
	\draw (C1) -- (C3) -- (C5);
	\draw (C2) -- (C4) node[above right]{$~~~~~e,\mathcal{E}$};
	
	\draw[blue,dashed] (D1) -- (D2) -- (D3);
	
	\draw[blue,dashed] (D1) -- ++($(0.5*\step:\sqrtthree*\r/3.0)$);
	\draw[blue,dashed] (D1) -- ++($(-1.5*\step:\sqrtthree*\r/3.0)$);
	
	\draw[blue,dashed] (D2) -- ++($(1.5*\step:\sqrtthree*\r/3.0)$) node[below right]{$\omega,H$};
	
	\draw[blue,dashed] (D3) -- ++($(2.5*\step:\sqrtthree*\r/3.0)$);
	\draw[blue,dashed] (D3) -- ++($(-1.5*\step:\sqrtthree*\r/3.0)$);
	
	\draw (C1) node {$\bullet$};
	\draw (C2) node {$\bullet$};
	\draw (C3) node {$\bullet$};
	\draw (C4) node {$\bullet$};
	\draw (C5) node {$\bullet$};
	
	\draw[blue] (D1) node {$\bullet$};
	\draw[blue] (D2) node {$\bullet$};
	\draw[blue] (D3) node {$\bullet$};
	\end{tikzpicture}

	\caption{The support graph $\Gamma$ for our holonomy functions is represented in blue here. A natural dual graph, in black can be constructed. The edges of this dual graph natural carry the momenta.}
	\label{fig:operatorsupport}
	
\end{figure}
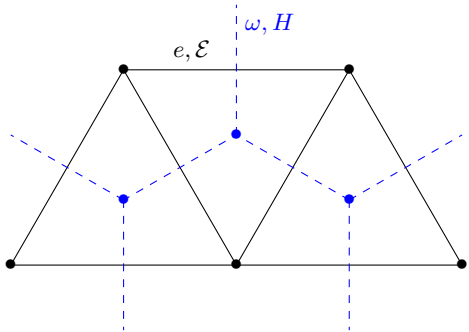

Let's now consider a graph $\Gamma$, that is a set of points $V$ linked by non-crossing paths\footnote{"non-crossing" can be understood as "every crossing has to be counted in the set of points $V$".} from a set $E$, embedded into $\Sigma$ the spatial manifold. To each of the paths in $E$, we can associate the holonomy along it. These holonomies, as functions over the connection, inherit a natural Poisson structure, which is trivial (as the connection commutes with itself). To construct the conjugate momenta, we should consider the graph $\tilde{\Gamma}$, dual to the original graph $\Gamma$. It is constituted by a set of path $\tilde{E}$ and a set of points $\tilde{V}$ with the following rule: for each path in $E$, there is crossing path in $\tilde{E}$ and for each plaquette formed by $\Gamma$, there is a point in $\tilde{V}$ which sits in the plaquette as illustrated on figure \ref{fig:operatorsupport}. The integrated triad along the paths from $\tilde{E}$ are the natural momenta. It can be shown \cite{Freidel:2010aq, Thiemann:2000bv} that, due to the parallel transport, they inherit a non-trivial Poisson structure from the continuum theory, which is:
\begin{equation}
\{\mathcal{E}^I_{\gamma},\mathcal{E}^J_{\gamma}\} = \epsilon^{IJ}_{~~K} \mathcal{E}^K_{\gamma}.
\end{equation}
As long as the paths do not cross, we can safely consider that the integrated triads commute if they are evaluated on different paths.

We should now discretize the constraints. As mentioned earlier, this is possible because the theory is topological. In essence quantizing a discrete subsystem correspond to solving nearly all the theory and keeping a small amount of degrees of freedom. As a consequence, the constraints need only be expressed for these remaining degrees of freedom. The Gauß constraints, in this regard, need only be expressed at the vertices in $V$. This is because the integrated variables are gauge-invariant nearly everywhere and depend on the gauge only at their start or end points, that is at the points in $V$. So, in effect, the Gauß constraints have been solved nearly everywhere by the the discretization. At these points, the natural discretization is:
\begin{equation}
\sum_{e/s(e) = v} \mathcal{E}^I_e - \sum_{e/t(e)=v} H_e^{-1} \triangleright \mathcal{E}^I_e = 0.
\end{equation}
Similarly, the flatness condition can be reproduced in the discrete setting by imposing all holonomies around closed loops $\ell$ of $\Gamma$ to be trivial:
\begin{equation}
\prod_{e \in \ell} H_e = \mathbb{I}.
\end{equation}
This concludes the standard analysis.

Let's conclude this section with a word on the quantization of the model, eventhough we will come back to it in the second part of the paper. As advertised so far, we will restrict to the discrete observables. This means, we will only represent holonomies along path of $\Gamma$ and their conjugated momenta. The representation is quite straightforward: we start with wave-functions over the configuration space which is simply $\# E$ copies of the gauge group $\mathrm{SU}(2)$. The operators associated to holonomies are represented by multiplication and the conjugated triad operator is represented by derivation. The constraints are then the natural quantization of the previous discrete expressions. It should be noted however that the Gauß constraints, because it is linear in the momentum, can be understood as the transformation of the configuration space and can be quantized in this manner. The solution is unique, though it is not the kinematical Hilbert space. It can be understood as a distribution over our Hilbert space and is the Dirac delta on the configuration space. We shall now turn to our problem.

\subsection{Simplicity constraints in 3d}

In this section, we will detail our new Hamiltonian analysis, leading to the extended phase space. As we have done in the first part, we start from the following action (using Planck units):
\begin{equation}
S[e,\omega] = \frac{1}{2} \int_\mathcal{M} \epsilon_{IJK} e^I \wedge F^{JK}[\omega].
\end{equation}
where $e$ is the triad and $\omega$ is a $\mathrm{SU}(2)$ connection. As before, $F$ represents the curvature of $\omega$ and reads:
\begin{equation}
F^{IJ}_{ab} = \partial_a \omega^{IJ}_{b} - \partial_b \omega^{IJ}_{a} + \omega^{IK}_{a} \omega^{LJ}_{b} \eta_{KL} - \omega^{IK}_{b} \omega^{LJ}_{a}\eta_{KL},
\end{equation}
and, in a more explicit manner, the action reads:
\begin{equation}
S[e,\omega] = \frac{1}{2} \int_\mathbb{R} dt \int_\Sigma d^2 \sigma \left(\epsilon^{abc} \epsilon_{IJK} e^I_a \left[\partial_b \omega^{IJ}_{c} - \partial_c \omega^{IJ}_{b} + \omega^{IK}_{b} \omega^{LJ}_{c} \eta_{KL} - \omega^{IK}_{c} \omega^{LJ}_{b}\eta_{KL}\right]\right).
\end{equation}
The goal is now to develop the Hamiltonian theory, but instead of considering the variable $\omega$ and $e$ as its conjugate momentum, we will consider $\omega$ and $e$ to be independent variables, with their respective momenta $B$ and $X$. Of course, this will lead to additional constraints which will turn out to be second class.

We can start by differentiating the Lagrangian with respect to time derivatives in order to get the expressions for the momenta. The Lagrangian $L$ is defined by:
\begin{equation}
S[e,\omega] = \int_\mathbb{R} dt L.
\end{equation}
This gives for the momenta conjugated to the spatial part of $e$ and $\omega$:
\begin{equation}
\begin{array}{rcl}
X^a_I(\sigma) &\equiv& \frac{\delta L}{\delta \dot{e}_a^I(\sigma)} = 0, \\
B^{a}_{IJ}(\sigma) &\equiv& \frac{\delta L}{\delta \dot{\omega}_{a}^{IJ}(\sigma)} = \epsilon_{KIJ} \epsilon^{ab} e^K_b(\sigma),
\end{array}
\end{equation}
where $\epsilon^{ab}$ is a totally antisymmetric tensor running only on spatial indices. We notice right away that the expressions cannot be inverted. The momenta associated to the time components simply give:
\begin{equation}
\begin{array}{rcl}
X^0_I(\sigma) &\equiv& \frac{\delta L}{\delta \dot{e}_0^I(\sigma)} = 0, \\
B^{0}_{IJ}(\sigma) &\equiv& \frac{\delta L}{\delta \dot{\omega}_{0}^{IJ}(\sigma)} = 0.
\end{array}
\end{equation}
Note that the system is maximally singular: no momenta can be expressed from the time derivatives of some field.

Let's derive the Hamiltonian for this formulation. The expression for the Hamiltonian $H$ is:
\begin{equation}
H = \int_\Sigma d^2 \sigma \left(X^0_I(\sigma) \dot{e}^I_0(\sigma) + \frac{1}{2}B^0_{IJ}(\sigma) \dot{\omega}_0^{IJ}(\sigma) + X^a_I(\sigma) \dot{e}^I_a(\sigma) + \frac{1}{2}B^{a}_{IJ}(\sigma) \dot{\omega}_{a}^{IJ}(\sigma)\right) - L.
\end{equation}
It is important to note that the Latin indices here run only on the spatial indices. Note also the factor $1/2$ which compensates the overrunning sum which exists because of the antisymmetry. Now, because the formulas for the momenta cannot be inverted, the time derivatives $\dot{e}$ and $\dot{\omega}$ will act as Lagrange multiplier. The goal is therefore to organize the terms in $L$ to whether they correspond to space or time indices. We find:
\begin{equation}
H = \int_\Sigma d^2 \sigma \left(\dot{e}^I_0 \Pi_I + \frac{1}{2} \dot{\omega}^{IJ}_0 \Xi_{IJ} + \dot{e}^I_a C^a_I + \frac{1}{2} \dot{\omega}^{IJ}_{a} S^{a}_{IJ} + e_0^I D_I + \frac{1}{2} \omega_{0}^{IJ} G_{IJ} \right),
\end{equation}
where $\sigma$ dependencies have been omitted for readability and:
\begin{eqnarray}
\Pi_I &=& X^0_I, \\
\Xi_{IJ} &=& B^0_{IJ}, \\
C^a_I &=& X^a_I, \\
S^{a}_{IJ} &=& B^{a}_{IJ} - \epsilon^{ab} \epsilon_{KIJ} e_b^K, \\
D_I &=& -\epsilon^{ab} \epsilon_{IJK} \left[ \partial_a \omega^{JK}_{b} - \partial_b \omega^{JK}_{a} + \omega^{JL}_{a} \omega^{MK}_{b} \eta_{LM} - \omega^{JL}_{b} \omega^{MK}_{a}\eta_{LM} \right], \\
G_{IJ} &=& - \epsilon^{ab} \epsilon_{KIJ} \partial_a e^K_b + \epsilon^{ab} \epsilon_{KIL} e_a^K\omega_{b}^{ML}\eta_{MJ} - \epsilon^{ab} \epsilon_{KLJ} e_a^K\omega_{b}^{ML}\eta_{MI}.
\end{eqnarray}
The primary constraints appear with the dotted factors. We can now apply the Dirac algorithm to devise all the constraints. It should be noted as this point however that the $e_0^I$ and $\omega_0^{IJ}$ factors are not Lagrange multiplier but true degrees of freedom. At least at this point, $D_I$ and $G_{IJ}$ do not appear as constraints.

The Dirac algorithm only needs two passes. Let's quickly describe them. $\Pi_I$, $\Xi_{IJ}$, $C^a_I$ and $S^a_{IJ}$ are primary constraints. In the first pass, we will check if their conservation implies secondary constraints. It is straightforward to check that $D_I = 0$ is implied by the conservation of $\Pi_I$ and that $G_{IJ} = 0$ is forced by the conservation of $\Xi_{IJ}$. The other constraints, $C^a_I$ and $S^a_{IJ}$, do not imply new constraints though. First, $C^a_I$ does not commute with $S^a_{IJ}$. This means that the conservation of $C^a_I$ will lead to equations fixing $\dot{\omega}^{IJ}_a$. Similarly the conservation of $S^a_{IJ}$ will lead to equations fixing $\dot{e}^I_a$. The equations are trivially invertible and therefore there is no other constraint in the first pass. In the second pass, we do the same analysis but with all the constraints $\Pi_I$, $\Xi_{IJ}$, $C^a_I$, $S^a_{IJ}$, $D_I$ and $G_{IJ}$. The analysis goes similarly. The conservation of $D_I$ and $G_{IJ}$ do not involve new constraints as it can be checked that the equations they imply are automatically solved if $C^a_I$ and $S^a_{IJ}$ are conserved. Therefore, after the second pass, the algorithm finishes and they are no tertiary constraints.  This system is obviously second class (they are some second class constraints). The next question is the classification of the constraints into first and second class.

To find the first-class constraints, we look for linear combinations of the previous constraints which commute with all the constraints. We can quickly find that $\Pi_I$ and $\Xi_{IJ}$ commute with all the other constraints. They are therefore first class. We will now look for the Gauss and the diffeomorphism constraints.

The following linear combination of the constraints is a good candidate for the Gauss constraint:
\begin{equation}
\mathcal{G}_{IJ} \equiv \partial_a S^a_{IJ} - \omega_{a}^{KL} S^{a}_{LJ} \eta_{IK} + \omega_{a}^{KL} S^{a}_{IL} \eta_{JK} - G_{IJ} + X^a_I e^K_a \eta_{KJ} - X^a_J e_a^K \eta_{KI}.
\end{equation}
Explicitly, it reads:
\begin{equation}
\mathcal{G}_{IJ} = \partial_a B^{a}_{IJ} - \omega_{a}^{KL} B^{a}_{LJ} \eta_{IK} + \omega_{a}^{KL} B^{a}_{IL} \eta_{JK} + X^a_I e^K_a \eta_{KJ} - X^a_J e_a^K \eta_{KI}.
\end{equation}
It is easy to check that these constraints have the correct algebra:
\begin{equation}
\{\mathcal{G}_{IJ},\mathcal{G}_{KL}\} = \eta_{IK} \mathcal{G}_{JL} - \eta_{IL} \mathcal{G}_{JK} - \eta_{JK}\mathcal{G}_{IL} + \eta_{JL} \mathcal{G}_{IK}.
\end{equation}
It is also straightforward to check that these constraints commute with all the other (on-shell). The last part in $X$ in particular is needed so that the constraint commute with $S^a_{IJ}$.

We can proceed in a similar fashion for the flatness condition. We consider the following linear combination of constraints:
\begin{equation}
\mathcal{D}_I \equiv D_I - \nabla_a X^a_I.
\end{equation}
It is easy to check that it commutes with all the other constraints on-shell.

Therefore, the initially found system of constraints is equivalent to a system governed by the following first class constraints:
\begin{eqnarray}
\mathcal{G}_{IJ} &=& \nabla_a B^{a}_{IJ} + X^a_I e^K_a \eta_{JK} - \eta_{IK} X^a_J e_a^K, \\
\mathcal{D}_I &=& \epsilon^{ab} \epsilon_{IJK}  F^{JK}_{ab} - \nabla_a X^a_I.
\end{eqnarray}
with the following second class constraints:
\begin{eqnarray}
C^a_I &=& X^a_I, \label{eq:momenta} \\
S^{a}_{IJ} &=& B^{a}_{IJ} - \epsilon^{ab} \epsilon_{KIJ} e_b^K. \label{eq:simplicity}
\end{eqnarray}

The choice of the second class constraints is somewhat arbitrary: it just has to be linear combinations of the constraints which are linearly independent from the first class constraints and constitute with them a complete basis. In particular, this implies that we can add any combination of the first class constraints to the second class ones. Our choice appears to be natural though as the corresponding constraints come out as primary constraints in our analysis. This concludes the Hamiltonian analysis.

We should note at this point that the analysis was straightforward and actually followed the usual Dirac procedure. It is also easy to check that the result reduces to the standard analysis when we explicitly solve the second class constraints. In this regard, this means that the usual analysis really is a quick hand way of doing the previous one followed by the explicit solution of the second class constraints. Our goal now will be quantize the theory without solving the second class constraints first. This will however first require that we quantize the system we are considering.

\subsection{Discrete setting}

As we mentioned earlier, the standard method in Loop Quantum Gravity to get a kinematical Hilbert space is to use a projective limit construction. This is however very hard to do in our case because of the introduction of the second class constraints. This is one of the reasons we considered a simplified 3d quantum gravity model. Indeed, 3d quantum gravity is topological (it doesn't have local degrees of freedom) and as such, the full model can be captured by a sufficiently refined discretization. The discretization should be considered as a subsystem. We will be looking for natural discrete observables on which the constraints can act without knowing the rest of the system, which is implicitly assumed to be in the physical state. This means in particular that a good strategy is to look for integrated variables such that the Poisson brackets between them is closed.

Let's start with the standard variables. The connection $\omega$, being a connection, has a natural integrated version through its holonomies. If we want to consider gauge invariant quantities, it is natural to consider conjugation invariant function over closed holonomies. But in general, we can associate an observable to a collection of $N$ paths embedded in $\Sigma$ and a function mapping $\mathrm{SU(2)}^N$ into $\mathbb{C}$. For each path $\gamma_i$ (parametrised by $s \in [0,1]$), we can define the holonomy $H_{\gamma_i}$. It is defined by:
\begin{equation}
H_{\gamma_i} = \tilde{H}_i(1),\quad \tilde{H}_i(0) = \id,\quad \frac{\mathrm{d}\tilde{H}_i}{\mathrm{d}s}(s) = \mathrm{i} \tilde{H}(s) \omega^{IJ}_a(\gamma_i(s)) J_{IJ} \dot{\gamma_i}^a .
\end{equation}
where $J_{IJ}$ are the generators of the $\mathrm{SU}(2)$ group. We can then have the observable $f(H_{\gamma_1},...H_{\gamma_N})$.

Similarly, the triad $e$ is a $\mathbb{R}^3$ valued one-form. It is therefore natural to consider its integration along one-dimensional paths. This integration is quite straightforward and reads:
\begin{equation}
\mathcal{I}(e)^I_{\gamma} = \int_\gamma (H_{\gamma_x} \triangleright e_a)^I \mathrm{d}x^a ,
\end{equation}
where $\gamma_x$ is the partial path, ending at the point $x$ of integration. As for the standard analysis, an important note should be made there. The partial path $\gamma_x$ does not have to start at the same point as the full path $\gamma$. It is conceivable to use path that are parallel transported further. For instance, we might consider a path $\gamma'$ starting at some point $y$ and ending on the starting point of $\gamma$. Then the following integrated quantity is perfectly reasonable:
\begin{equation}
\mathcal{I}(e)^I_{\gamma,\gamma'} = \int_\gamma (H_{\gamma'\cup\gamma_x} \triangleright e_a)^I \mathrm{d}x^a .
\end{equation}
This corresponds to the parallel transport of the previous quantity along $\gamma'$. This option will be interesting when considering conjugated momenta.

As we did earlier, we can consider observables using a function mapping $\mathbb{R}^3$ into $\mathbb{C}$ mimicking the usual construction of holonomies. In a sense, these integrated version are $\mathbb{R}^3$ holonomies. As usual holonomies, they are gauge invariant nearly everywhere and gauge covariant at the end points of $\gamma$. The gauge group in question however is $\mathbb{R}^3$ which is not a fundamental gauge group for our theory (which is only $\mathrm{SU}(2)$ invariant)\footnote{We are not considering here the Chern-Simons formulation uncovered by Witten \cite{Witten:1988hc} for which the gauge group is extended to $\ISU(2)$. This is very peculiar to 3d and as such not very interesting in the perspective we have.}. The action of the usual $\mathrm{SU}(2)$ gauge group is still natural: the integrated quantity transforms covariantly by the gauge transformation at $s(\gamma)$ (the start point of the path). Note that there is a nice natural geometrical interpretation for infinitesimal closed loops as it evaluates to the torsion of $e$ with respect to $\omega$.

We can now consider the momenta. Both momenta are densitized vector field valued in a Lie algebra. $X$ is valued in the trivial Lie algebra $\mathbb{R}^3$ and $B$ is valued in the $\mathfrak{su}(2)$ Lie-algebra which is isomorphic, as a vector space, to $\mathbb{R}^3$. In 3d quantum gravity, there is a natural correspondence between densitized vector field and one-forms. This means that the momenta will also be integrated along lines. In the context of 3d quantum gravity, the fact that they are densitized vector field is important with respect to the simplectic structure: their conjugation is between a variable and the corresponding momentum evaluated in orthogonal directions.

But for now, the integrated versions are defined in ways very similar to the integrated triad, using now the Levi-Civita tensor to get the duality between densitized vectors and one-form. Concretely, we have:
\begin{equation}
\mathcal{I}(X)_{I,\gamma} = \int_\gamma (H_{\gamma_x} \triangleright X^a)_I \epsilon_{ab} \mathrm{d}x^b ,
\end{equation}
and:
\begin{equation}
\mathcal{I}(B)_{IJ,\gamma} = \int_\gamma (H_{\gamma_x} \triangleright B^a)_{IJ} \epsilon_{ab} \mathrm{d}x^b .
\end{equation}
As mentioned previously, we could consider further parallel transport and this will indeed be important in what follows.

Note at this point that we have not defined natural integration for the variables $e_0^I$ and $\omega_0^{IJ}$. If we had to do so, they should actually evaluate on points (they are scalar quantities with respect to spatial diffeomorphism). But because of the (first class) constraints $\Pi_I$ and $\Xi^{IJ}$, we know that the quantum wave-function will not depend on these variables, nor will the classical theory. We can just as well dismiss them altogether.

Let's now consider what integration path we should be using, with what potential further parallel transport. Our first clue comes from the usual construction: one starts with the holonomies integrated along the lines of a specific graph and then the momenta are integrated over dual lines. In our case however, this is rather more subtle as we have four different types of quantity to integrate and only pairs of line to work with. A natural choice would be to start with $\omega$ and $e$ integrated on the lines of a graph $\Gamma$. The conjugated momenta $X$ and $B$ would then be integrated along the dual paths. The theory seems however to favor another possibility. Indeed, as we look to discretize the system, we also look to be able to write down the various constraints. The constraint $B=e$ in particular seems to tell us that $B$ and $e$ should be integrated along the same path. This leads to the following choice: we will integrate $\omega$ and $X$ and the same path and the momenta $B$ and the triad $e$ will be integrated on a dual path. As is standard in 3d quantum gravity, the momentum will be transported to the same starting as the configuration variables, as was already illustrated on figure \ref{fig:momentumintegration}. This is needed for a better covariance of the Poisson brackets. There is however a substantial problem with the construction we just offered: the Poisson brackets are not closed. It is possible to check for instance that the Poisson brackets between $\mathcal{I}(B)$ and $\mathcal{I}(e)$ involve quantities integrated over partial path from the graph.

This leads us to finding natural integration with a closed Poisson algebra. A straightforward possibility is to try and form an $\mathrm{ISU}(2)$, or $\mathrm{T}^*\mathrm{ISU}(2)$, algebra. This is possible using the following integrated variables:
\begin{equation}
\left\{
\begin{array}{rcl}
I(\omega)_\gamma &=& H_\gamma, \\
I(X)_\gamma &=& \mathcal{I}(X)_\gamma, \\
I(B)_\gamma &=& H_\gamma' \triangleright \mathcal{I}(B)_{\tilde{\gamma}} + \int_{\tilde{\gamma}} \mathcal{I}(X)_{\gamma ' \cup \tilde{\gamma}_x} \times (H_ {\gamma '} \triangleright \mathcal{I}(e)_{\tilde{\gamma}_x}) \mathrm{d}x, \\
I(e)_\gamma &=& H_\gamma' \triangleright \mathcal{I}(e)_{\tilde{\gamma} }.
\end{array}
\right.
\end{equation}
In these definitions, $\gamma$ is an edge from the graph $\Gamma$. $\tilde{\gamma}$ is the corresponding dual edge. A subscripted edge, like $\gamma_x$, is the edge but stopped at the point $x$. The primed edge $\gamma'$ is the path starting as $\gamma$ until the crossing point $y$ with its dual. At $y$, the primed edge continues as $\tilde{\gamma}_y^{-1}$, that is the reverse path from $y$ to the starting point of $\tilde{\gamma}$. The cross $\times$ is simply the vector cross product. The definition is rather straightforward except for the term $I(B)$. It is the integrated $B$ momentum parallel transported but with an additional term. This additional term generates the rotation of $X$ and $e$ and is needed for the algebra to close. This leads to the closed algebra:
\begin{equation}
\left\{
\begin{array}{rcl}
\{I(B)_{\gamma_i}, I(\omega)_{\gamma_j}\} &=& \delta_{ij} \sigma^i I(\omega)_{\gamma_j}, \\
\{I(B)_{\gamma_i}, I(B)_{\gamma_j}\} &=& \delta_{ij} \epsilon I(B)_{\gamma_j}, \\
\{I(B)_{\gamma_i}, I(e)_{\gamma_j}\} &=& \delta_{ij} \epsilon I(e)_{\gamma_j}, \\
\{I(B)_{\gamma_i}, I(X)_{\gamma_j}\} &=& \delta_{ij} \epsilon I(X)_{\gamma_j}, \\
\{I(X)_{\gamma_i}, I(e)_{\gamma_j}\} &=& \delta_{ij}, \\
\end{array}
\right.
\end{equation}
all other brackets being zero. The final point to have a discrete theory is to express the constraints. It is rather similar to the expression of the constraints in the usual analysis. The Gauß constraint is expressed at every node of the graph. The sum of the $I(B)$ terms around a vertex correctly generates the rotation as can be seen from the previous algebra. The flatness constraints is a bit more technical but comes down to expressing the flatness of the connection $\omega + X$. We will detail the full expression when studying the quantum theory.

\section{Gauge unfixing and quantum theory}

\subsection{Second class constraints, gauge unfixing}

Before actually quantizing the classical theory we just presented, we should consider how to deal with the second class constraints. Indeed, contrary to what can be done for first class constraints, it is impossible to impose (simultaneously) second class constraints as operator constraints in the quantum theory.

There are several techniques to tackle this problem in a quantum theory. A usual one is to use the Gupta-Bleuler method. This is however difficult to implement in a diffeomorphism invariant theory. To make things more precise, let's imagine that the constraints simply read $\phi = 0$ and $\Pi_\phi = 0$ with the bracket $\{\phi,\Pi_\phi\} = \delta$. The Gupta-Bleuler method consist in implementing a holomorphic constraint $\phi + \mathrm{i}\Pi_\phi = 0$ which, because it does not Poisson-commute with its conjugate, is not diagonalizable in an orthogonal basis. The solutions are therefore states which are spread (in our example, they are even coherent states) and carry this way the \textit{soft} implementation of the constraints. The problem is that $\phi$ and $\Pi_\phi$ are not of the same density. With a given background metric, it is possible to define the sum by using the hodge dual but in a diffeomorphism invariant setting, this definition is no longer possible. As we are working on a discrete lattice, it is in fact possible, in principle, to use the Gupta-Bleuler method. It will however make the formulation of a continuum theory rather hard.

Here, to test our quantum model, we will follow \cite{Bodendorfer:2011nv} and use gauge unfixing. The basic idea is that first class constraints when gauge-fixed lead to sets of second class constraints. In a similar manner, it is possible to unfix gauge from second class constraints and get a set of first class constraints out of it. In our case, it is made very simple by the following fact: both second class constraints commute with all the first class constraints without the need of imposing other constraints. So, when $X \simeq 0$, $X$ commutes with all the (first class) constraints and similarly when $B-e \simeq 0$, $B-e$ commutes with all the (first class) constraints. Each constraint can therefore be thought of as a first class constraint gauge fixed by the other one. We are then led to a choice of choosing which constraint to keep.

Our choice will be guided by simplicity. In the discrete setting, we have no natural integrated observable corresponding to $B-e$, but we have one for integrated $X$. We will therefore quantize the following sets of constraints:
\begin{eqnarray}
\mathcal{G}_{IJ} &=& \nabla_a B^{a}_{IJ} + X^a_I e^K_a \eta_{JK} - \eta_{IK} X^a_J e_a^K, \\
\mathcal{D}_I &=& \epsilon^{ab} \epsilon_{IJK}  F^{JK}_{ab} - \nabla_a X^a_I, \\
C^a_I &=& X^a_I.
\end{eqnarray}
We will consider the corresponding quantum theory.

We should note before continuing that gauge unfixing is not the only option. One of the major ways to deal with second class constraints is rather the Dirac algorithm and the introduction of the Dirac brackets. In our case however, this would not be useful. The reason is that introducing the Dirac brackets will lead to a theory which is more or less completely equivalent to the standard approach. Indeed, it can be checked that:
\begin{eqnarray}
\{\omega_a^{IJ} + \epsilon_{ac} \epsilon^{IJM} X_M^c, \omega_b^{KL} + \epsilon_{bd} \epsilon^{KLN} X_N^d\}_D &=& 0 \\
\{\omega_a^{IJ} + \epsilon_{ac} \epsilon^{IJM} X_M^c, B^b_{KL}\}_D &=& \delta_a^b (\delta^I_K \delta^J_L - \delta^I_L \delta^J_K) \\
\{\omega_a^{IJ} + \epsilon_{ac} \epsilon^{IJM} X_M^c, X^b_K\}_D &=& 0 \\
\{\omega_a^{IJ} + \epsilon_{ac} \epsilon^{IJM} X_M^c, B^b_{KL} - \epsilon^{bd} \epsilon_{IJN} e^N_d\}_D &=& 0 \\
\{B^a_{IJ}, B^b_{KL}\}_D &=& 0 \\
\{B^a_{IJ}, X^b_K\}_D &=& 0 \\
\{B^a_{IJ}, B^b_{KL} - \epsilon^{bd} \epsilon_{IJN} e^N_d\}_D &=& 0 \\
\{X^a_I, X^b_J\}_D &=& 0 \\
\{X^a_I, B^b_{KL} - \epsilon^{bd} \epsilon_{IJN} e^N_d\}_D &=& 0 \\
\{B^a_{IJ} - \epsilon^{ac} \epsilon_{IJM} e^M_c, B^b_{KL} - \epsilon^{bd} \epsilon_{IJN} e^N_d\}_D &=& 0
\end{eqnarray}
This shows that once the second class constraints are imposed, all that will remain is the two conjugated variables $\omega_a^{IJ} + \epsilon_{ac} \epsilon^{IJM} X_M^c$ and $B^a_{IJ}$. We can then re-express all the other constraints using them and find that this is exactly the standard way of presenting 3d gravity. Because we wanted to study the imposition of the second class constraints, this is not very interesting. It is worth noting however that this is the case because the simplicity constraints are trivial. In higher dimensions, for instance in 3+1d, we do not expect the Dirac brackets to be so simple. It might therefore be an interesting way of treating the fuller problem.

\subsection{Hilbert space and observables}

For each variable we defined earlier, we will need a quantum operator acting on some Hilbert space. We will first concentrate on the definition of the Hilbert space. Let's recall how the usual Hilbert space for 3d (discrete) quantum gravity. We are given a graph $\Gamma$ embedded in the space manifold $\Sigma$. Each edge carries a copy of the Hilbert space $\mathrm{L}^2(\mathrm{SU}(2),\mathrm{d}h)$ of square integrable functions over $\mathrm{SU}(2)$ equipped with the natural (left and right) Haar measure. We denote $\mathcal{H}_e$ the copy associated to an edge $e$. The full Hilbert space is the tensor product over all edges:
\begin{equation}
\mathcal{H} = \bigotimes_{e \in E} \mathcal{H}_e .
\end{equation}
The holonomy along the edge $e$ simply acts by multiplication on the corresponding Hilbert space. The integrated conjugated momentum on a crossing line acts by derivation on the Hilbert space corresponding to the crossed edge.

Let's move on to our construction. The main choice boils down to the choice of Hilbert space for a single edge. As we mentioned earlier, there are various possible discretizations for our model. As will become clear when we will study the constraints, one choice stands out: we should define a natural Poincaré holonomy, constructed out of $\omega$ and $X$. It should be natural to consider $\mathrm{L}^2(\mathrm{ISU}(2),\mathrm{d}\tilde{h})$ where $\mathrm{d}\tilde{h}$ is the natural Haar measure on the Poincaré group. Now, this is problematic, \textit{a priori} at least on two accounts. First, the (3d euclidean) Poincaré group is not compact. Therefore talking about \textit{the} Haar measure should be frowned upon. There are two obstructions for uniqueness. First it will not be possible to normalize the measure, leading to a family of Haar measures. Second, because the group is not compact, left and right invariance might not lead to the same measure. There is a second problem with choosing such a measure on a non-compact space: it will be extremely difficult to define a projective limit from there. Let's deal which each point separately.

Let's concentrate on the problem of left and right invariance. We recall first the multiplication of the group $\mathrm{ISU}(2) \simeq \mathbb{R}^3 \rtimes \mathrm{SU}(2)$. Let $(\vec{T},g)$ and $(\vec{T}',g')$ be two elements of $\mathrm{ISU}(2)$ written down in the natural decomposition following the previous semidirect product. Then we have:
\begin{equation}
(\vec{T},g)\star(\vec{T}',g') = (\vec{T} + g \triangleright \vec{T}', gg') ,
\end{equation}
where $\triangleright$ is the natural action of $\mathrm{SU}(2)$ onto $\mathbb{R}^3$. Since $\mathrm{ISU}(2)$ is homeomorphic, as a topological space, to $\mathbb{R}^3 \times \mathrm{SU}(2)$, let's define a natural measure on this space and check that it is indeed left and right invariant. We consider the product measure $\mathrm{d}^3T \mathrm{d}h$ where $\mathrm{d}^3T$ is the usual Haar measure on $\mathbb{R}^3$ (which is also the Lebesgue measure) normalized so that the unit cube has volume $1$, and $\mathrm{d}h$ is the Haar measure on $\mathrm{SU}(2)$ normalized so that the whole space is of measure $1$. An important property of the Lebesgue measure is that it is spherically symmetric. Therefore, as a Haar measure, it is invariant under translation. But as a spherically symmetric measure, it is invariant under the action of $\mathrm{SU}(2)$. It is easy to check that this measure is invariant under the left action of $\mathrm{ISU}(2)$ group. Indeed, formally we have:
\begin{equation}
\mathrm{d}^3(\vec{T} + g \triangleright \vec{T}') \mathrm{d}(gg') = \mathrm{d}^3(g \triangleright \vec{T}') \mathrm{d}(g') = \mathrm{d}^3(\vec{T}') \mathrm{d}(g').
\end{equation}
The first equality comes from the left invariance of both measures. The last equality comes from the invariance of the Lebesgue measure under the action of $\mathrm{SU}(2)$. We can check in a similar manner that the measure is right-invariant:
\begin{equation}
\mathrm{d}^3(\vec{T} + g \triangleright \vec{T}') \mathrm{d}(gg') = \mathrm{d}^3(\vec{T}) \mathrm{d}(g).
\end{equation}
Here, only the right invariance of each measure is needed. Because the measure we constructed satisfies all the nice properties of the Haar measure on $\mathrm{ISU}(2)$, we have proven that we have a both left and right invariant measure on the group. This means the group is unimodular. Note that this depends heavily on the existence of a left and right Haar measure on the proper compact subgroup $\mathrm{SU}(2)$. The proof therefore generalizes to higher dimensions as long as we consider euclidean signatures. The Lorentzian case should be handled with more care. Regarding the normalization problem, we have settled it in the previous discussion by considering natural normalization for the Haar measures on $\mathbb{R}^3$ and $\mathrm{SU}(2)$. In what follows, we will consider this normalization.

The problem of the projective limit is rather more difficult to treat. There are various possibilities however. One of the easiest way to treat the problem is to choose a discrete topology\footnote{This is possible only because we are not looking for a Poincaré invariant theory and because the group is not simple. Such a construction for $\mathrm{SL}(2,\mathbb{C})$ via the Iwasawa decomposition for instance, will not give a correct projective limit. Indeed, the Iwasawa decomposition will need the selection of a privileged frame.} on $\mathbb{R}^3$. This will make impossible the definition of the integrated triads, except through their exponentials. The other natural possibility is to consider the Bohr compactification of $\mathbb{R}^3$. It is difficult to write the action of $\mathrm{SU}(2)$ on the Bohr compactification but it is possible to write the action on the dual. This corresponds therefore to write a theory with $\omega$ and $e$ operators acting by multiplication, where the $e$ operator spectrum is equipped with discrete topology. This will make impossible the definition of the canonical momenta $X$, except through their exponentials. Because $B$ and $e$ are compared in the constraint $B-e \simeq 0$, it should be more natural to have $e$ defined (as $B$ is naturally defined) but other possibilities could be argued in different settings. We will continue for now, in the discrete with our natural Haar measure associated to the natural topology on $\mathrm{ISU}(2)$.

Therefore, for each edge $e \in E$ of the graph $\Gamma$, we construct a copy of $\mathrm{L}^2(\mathrm{ISU}(2),\mathrm{d}\tilde{h})$ which we denote $\tilde{\mathcal{H}}_e$. The full Hilbert space is given by the tensor product:
\begin{equation}
\tilde{\mathcal{H}} = \bigotimes_{e\in E} \tilde{\mathcal{H}}_e .
\end{equation}
This space can be understood as the natural square integrable functions over $\# E$ copies of $\mathrm{ISU}(2)$. A wave-function in this space is of the form $\psi(\vec{T}_1,g_1,\vec{T}_2,g_2,...,\vec{T}_{\#E},g_{\#E})$. And the natural scalar product is:
\begin{equation}
\langle \psi | \phi \rangle = \int \overline{\psi(\vec{T}_1,g_1,\vec{T}_2,g_2,...,\vec{T}_{\#E},g_{\#E})} \phi(\vec{T}_1,g_1,\vec{T}_2,g_2,...,\vec{T}_{\# E},g_{\# E}) \prod_{i=1}^{\# E} \mathrm{d}^3 \vec{T}_i \mathrm{d}g_i .
\end{equation}
We have natural operators associated to each edge. The operators associated to $\omega$ and $X$ act by multiplication:
\begin{eqnarray}
\widehat{f(H_i)} \psi (\vec{T}_1,g_1,\vec{T}_2,g_2,...,\vec{T}_{\#E},g_{\#E}) &=& f(g_i) \psi(\vec{T}_1,g_1,\vec{T}_2,g_2,...,\vec{T}_{\#E},g_{\#E}) , \\
\widehat{I(X)_{I,i}} \psi(\vec{T}_1,g_1,\vec{T}_2,g_2,...,\vec{T}_{\#E},g_{\#E}) &=& T_{I,i} \psi(\vec{T}_1,g_1,\vec{T}_2,g_2,...,\vec{T}_{\#E},g_{\#E}) ,
\end{eqnarray}
where $f$ is any smooth function from $\mathrm{SU}(2)$ onto $\mathbb{C}$. The natural operators associated to $B$ and $e$ act by differentiation:
\begin{eqnarray}
\widehat{I(B)_{IJ,i}} \psi (\vec{T}_1,g_1,\vec{T}_2,g_2,...,\vec{T}_{\#E},g_{\#E}) &=& -\mathrm{i} \psi(\vec{T}_1,g_1,\vec{T}_2,g_2,...,\vec{T}_i,J_{IJ} g_i,...,\vec{T}_{\#E},g_{\#E}) \nonumber \\
&~& - i T_{[I,i} \frac{\partial \psi}{\partial T_{J],i}}(\vec{T}_1,g_1,\vec{T}_2,g_2,...,\vec{T}_{\#E},g_{\#E}), \\
\widehat{I(e)^I_{i}} \psi (\vec{T}_1,g_1,\vec{T}_2,g_2,...,\vec{T}_{\#E},g_{\#E}) &=& -\mathrm{i} \frac{\partial \psi}{\partial T_{I,i}}(\vec{T}_1,g_1,\vec{T}_2,g_2,...,\vec{T}_{\#E},g_{\#E}).
\end{eqnarray}
Here, the insertion of $J$ in the first equation correspond to the insertion of the corresponding generator of the $\mathfrak{su}(2)$ Lie-algebra in the Peter-Weyl summation (the Fourier transform). This is the natural differentiation on $\mathrm{SU}(2)$. The extra term generates the rotation of $e$ and $X$ and is just of the form $e \wedge X$.

The operators associated to the holonomies are unitary, while all the others are hermitian. This concludes the definition of our working Hilbert space with support on a fixed, given graph.

\subsection{Implementation of the constraints}

Let us discuss the implementation of the constraints. The Gauß constraints are the easiest to discretize. As they generated the $\mathrm{SU}(2)$ rotations, we just have to find their generators. In the usual standard 3d euclidean gravity, this is done at each vertex of the support graph, by summing all the incoming momenta and subtracting the outgoing momenta. This can be traced back to the Poisson algebra of the variables which reproduces the algebra of $\mathrm{T}^\star \mathrm{SU}(2)$. In our case, this is very similar, except, the Poisson algebra is the algebra of $\mathrm{T}^\star \mathrm{ISU}(2)$.

In practice, the classical constraint reads
\begin{equation}
\mathcal{G}_{IJ} = \nabla_a B^{a}_{IJ} + X^a_I e^K_a \eta_{JK} - \eta_{IK} X^a_J e_a^K .
\end{equation}
The first part $\nabla_a B^{a}_{IJ}$ is exactly the same as the usual case. The divergence operator corresponds to the summing around a vertex. We have extra terms corresponding to the rotation of $X$ and $e$. These however already appear in the discretization of our discretized integrated momenta $I(B)$. This leads us to the following Gauss constraint at a given vertex:
\begin{equation}
\widehat{\mathcal{G}_{IJ,v}} = \sum_{e \in E,~ s(e) = v} \widehat{I(B)_{IJ,e}} - \sum_{e \in E,~ t(e) = v} \widehat{H_e^{-1}} \triangleright \widehat{I(B)_{IJ,_e}}.
\end{equation}
The $\triangleright$ notation should be made more precise. It encodes the parallel transport needed for incoming momenta. Explicitly, it reads:
\begin{equation}
\widehat{H_e^{-1}} \triangleright \widehat{I(B)_{IJ,_e}} \psi (\vec{T}_1,g_1,\vec{T}_2,g_2,...,\vec{T}_{\#E},g_{\#E}) = \left(g_e^{-1}\right)_I^{\phantom{I}K} \left(g_e^{-1}\right)_J^{\phantom{J}L} \widehat{I(B)_{KL,e}} \psi (\vec{T}_1,g_1,\vec{T}_2,g_2,...,\vec{T}_{\#E},g_{\#E}) .
\end{equation}
It is easy to check that these constraints have the right algebra.

Let's now consider the flatness constraints. They read:
\begin{equation}
\mathcal{D}_I = \epsilon^{ab} \epsilon_{IJK}  F^{JK}_{ab} - \nabla_a X^a_I.
\end{equation}
These constraints are the reason we selected our integration eventhough we are not implementing the constraint $B-e$ at the quantum level. Indeed, we can see that we will need to integrate $\omega$ and $X$ on the same structure, which favours choosing $X$ as the translation part of the Poincaré connection.

We can use the natural discretized version of $F(\omega)$. This will be done by considering closed loops in the support graph and the corresponding holonomy operator in the fundamental representation around these loops. Similarly, we can discretize the divergence of $X$ by integrating the variable along a closed loop. This will lead to an $\mathbb{R}^3$ element which is not directly comparable to the integrated holonomy, but we can work around this by exponentiating the vector into an $\mathrm{SU}(2)$ matrix. Concretely, we define the following constraints: for any closed loop $\ell$ on our graph, we impose:
\begin{equation}
\widehat{\mathcal{D}_{\ell}} = \left(\prod_{e\in \ell} \widehat{H_e}\right) \left(\prod_{e \in \ell} \exp \left[ - \frac{\mathrm{i}}{2} \prod_{e' \in \ell}^{e} \widehat{H_{e'}} \triangleright \widehat{I(X)_{I,e}} \sigma^I \right]\right)^{-1} - \mathbb{I}.
\end{equation}
Here the products are ordered so that multiplication happens in order along the path. The operator is matrix valued and it written this way for readability but could of course be split into $4$ different operators. They are no ordering problem because the fundamental variables commute with each other. Note also the parallel transport in the exponential, which is cumbersome but needed for gauge covariance.

The last constraint is the gauge-unfixed version of the simplicity constraints which simply reads:
\begin{equation}
C^a_I = X^a_I.
\end{equation}
Because it is so simple, it can be implemented as a constraint on each link $e$ with:
\begin{equation}
\widehat{C_{I,e}} = I(X)_{I,e}.
\end{equation}
This concludes the definition of the quantum theory.

It is a remarkable fact that the previous construction is equivalent to the standard euclidean 3d quantum gravity. It can be checked explicitly by looking for the kernel of all the constraints. But more easily, it can be checked by verifying that the kernel of the (retained) simplicity constraint is indeed the usual kinematical Hilbert space and that the other constraints reduce to the standard ones.

Let's look for functions $\Psi$ inside our Hilbert space such that the constraints $\widehat{C_{I,e}} \psi = 0$ are satisfied. Let's note first that there is no such functions. Indeed, the eigenvectors of these constraints are Dirac deltas which cannot be normalized. This is however standard: first class constraints generically lead to non-normalizable eigenstates, which can however be equipped with a norm if one restrict to the kernel of the constraints. Note that this whole discussion can be avoided by just choosing a discrete measure of our translation part of our group element to start with, as was needed anyway for the projective limit.

This condition therefore reduces our Hilbert space to the standard kinematical space. More explicitly, we have:
\begin{equation}
\bigcap_{e \in \Gamma, I \in \{1,2,3\}} \ker \widehat{C_{I,e}} \simeq \bigoplus_{e \in \Gamma} \mathrm{L}^2(\mathrm{SU}(2)).
\end{equation}
The interesting bit is to see that on these bit, the other constraints reduce to the standard constraints.

For the Gauß constraints, this works because the point selected for $X$, namely $X=0$, is invariant under gauge transform. Therefore, the gauge transformation acts trivially on this part of the $\mathrm{ISU}(2)$ group and reproduces exactly the usual action on the $\mathrm{SU}(2)$ group.

For the flatness constraints, we can see this explicitly. We can explore the action of the flatness constraint on a solution of the constraints $\psi$. As,
\begin{equation}
\left(\prod_{e \in \ell} \exp \left[ - \frac{\mathrm{i}}{2} \prod_{e' \in \ell}^{e} \widehat{H_{e'}} \triangleright \widehat{I(X)_{I,e}} \sigma^I \right]\right)^{-1}\psi = \psi,
\end{equation}
since $\hat{X}$ has a trivial action. This implies:
\begin{equation}
\widehat{\mathcal{D}_{\ell}}\psi = \left(\prod_{e\in \ell} \widehat{H_e}\right) \psi - \mathbb{I}\psi.
\end{equation}
We find indeed the usual flatness constraint which proves the equivalence.

\section{Outlook}

\subsection{Regge geometries}

What all this work has proved so far is that indeed our model can be defined and is equivalent, at the classical as well as at the quantum level, to the usual quantum gravity construction. The interest of this model is however to be able to explore new methods, and not restrict to methods that have been investigated, at least partially, in the full theory \cite{Achour:2015xga, Achour:2014rja, Achour:2014eqa, Wilson-Ewing:2015lia, Wilson-Ewing:2015xaq, Thiemann:1994pj, Thiemann:1992jj}.

It is worth noting in this regard that extending the framework to 3+1d might lead to an interesting Regge-like representation for Loop Quantum Gravity. Indeed, as we hinted at in the introduction, work by Dittrich \textit{et al.} \cite{Dittrich:2012rj} might be used as an argument to say that the usual reduction of the simplicity constraints using the Ashtekar-Barbero variables is actually incomplete and should be reduced further with some Regge geometry condition or shape matching condition. Equivalently, one might try to implement the simplicity constraints at the quantum level in such a way that Regge geometries automatically come out. For this, our framework might be useful.

Indeed, going from 3 dimensions to 4 dimensions, the simplicity constraints won't be as trivial. If we still keep the tetrad and the independent spin connection, the simplicity constraints will read:
\begin{equation}
B = e \wedge e,\quad X = 0,
\end{equation}
noting that they are additional constraints that come from the full analysis, as is detailed in appendix \ref{app:constraints4d}. Actually enforcing these conditions might be quite hard and is not particularly straightforward if we want to keep the compatibility with the projective limit construction. We can however sketch what happens in a discrete, simplified case.

\begin{figure}[h!]
	\centering
	
	\begin{tikzpicture}[scale=1]
	\coordinate (O1) at (0,0,0);
	\coordinate (A1) at ($(O1) + (0,2.,0)$);
	\coordinate (B1) at ($(O1) + (0,-1,-1.732)$);
	\coordinate (C1) at ($(O1) + (0,-1,1.732)$);
	\coordinate (D1) at ($(O1) + (-4.,-1,0)$);

	\coordinate (O2) at (2.,0,0);
	\coordinate (A2) at ($(O2) + (0,2.,0)$);
	\coordinate (B2) at ($(O2) + (0,-1,-1.732)$);
	\coordinate (C2) at ($(O2) + (0,-1,1.732)$);
	\coordinate (D2) at ($(O2) + (4.,-1,0)$);
	
	\draw[blue] (A1) -- node[midway,right,red] {$e_2$} (B1);
	\draw[blue] (A1) -- node[midway,left,red] {$e_1$} (C1);
	\draw[blue] (A1) -- (D1);
	\draw[blue] (B1) -- node[midway,right,red] {$e_3$} (C1);
	\draw[blue] (C1) -- (D1);
	\draw[dashed,blue] (D1) --  (B1);

	\draw[dashed,blue] (A2) -- node[midway,right,red] {$e_2$} (B2);
	\draw[blue] (A2) -- node[midway,above left,red] {$e_1$} (C2);
	\draw[blue] (A2) -- (D2);
	\draw[dashed,blue] (B2) -- node[midway,right,red] {$e_3$} (C2);
	\draw[blue] (C2) -- (D2);
	\draw[dashed,blue] (D2) --  (B2);
	
	\draw[dotted,red] (A1) -- (A2);
	\draw[dotted,red] (B1) -- (B2);
	\draw[dotted,red] (C1) -- (C2);
	
	\draw[blue] (A1) node{$\bullet$};
	\draw[blue] (B1) node{$\bullet$};
	\draw[blue] (C1) node{$\bullet$};
	\draw[blue] (D1) node{$\bullet$};

	\draw[blue] (A2) node{$\bullet$};
	\draw[blue] (B2) node{$\bullet$};
	\draw[blue] (C2) node{$\bullet$};
	\draw[blue] (D2) node{$\bullet$};
	\end{tikzpicture}
	
	\caption{The usual twisted geometry framework does not include shape matching conditions for the reconstructed geometry. In the approach we suggest, the natural quantities associated to the tetrad $e$ are carried by the edges of the triangulation dual to the support graph. This means that a natural matching of them should imply shape matching conditions. This support the idea that further development of our technique might lead to a Regge-like representation for LQG.}
	\label{fig:regge}
\end{figure}
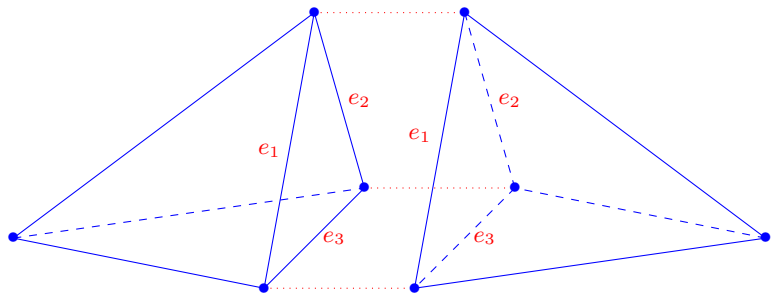

Let's consider only the condition $B = e \wedge e$. Ignoring the Hamiltonian structure, and concentrating on triangular faces, we can see that this condition should be discretized in the following way: for each edge of a face, we associated a vector $E$ (representing an integrated version of $e$) and for each face, we associated a bivector $I_B$ (representing an integrated version of $B$). If the sum of the integrated $E$ around a face is $0$ (which makes sense as it is a discretization of the no-torsion condition), then the simplicity constraint can be written as $I_B = E_1 \times E_2$, where the choice of the vectors $E_1$ and $E_2$ does not matter (thanks to the closure around faces) and the $\times$ is the cross-product from vectors to bi-vectors. The resulting theory should then give Regge-like geometry as the usual area matching is now supplemented by matching of the tetrads (see figure \ref{fig:regge}).

However, we have just glossed over many technical difficulties. As we mentioned already, it is very difficult to have a construction which is compatible with the projective limit and for which we can therefore give a natural continuum version. As we rapidly said, we also neglected the problem of the Hamiltonian structure. As in the 3d case, it is difficult to find a discretization which has a nice Hamiltonian structure. All this means that our previous argument is hand-waving at best. But it is still possible indeed, that the tetrad variables might help build a Regge-like quantum geometry. The argument at least suggest that keeping the variables $e$ can act as a guide in finding such a representation. This is why we have given the full Hamiltonian analysis in appendix \ref{app:constraints4d}, as a starting point for further inquiries.

\subsection{Self-dual variables}

Now, we explore a second possibility of these toy-models and one of their possible use for 4d quantum gravity: we think this model could be interesting in the study of self-dual variables. Let's explore this idea here.

In full 3+1d loop quantum gravity, one might want to use the self-dual Ashtekar variables. They have the very nice geometric property of being true pull-back from a space-time connection and, as such, simplify many expressions. Their geometric interpretation is also a desirable property when going to a quantum theory in order to avoid quantum anomalies. But in order to implement these variables in the quantum realm, one must also implement the second class constraints of the theory. This is indeed needed as the Ashtekar variables are in fact complex and encodes twice the needed amount of degrees of freedom.

The fact that the Ashtekar connection is complex turns out to be quite interesting from another perspective: the simplicity constraints map to reality constraints which can be interpreted as constraints on the scalar product for the kinematical Hilbert space (as it is partially implemented for Loop Quantum Cosmology and symmetry reduced models in \cite{Wilson-Ewing:2015lia, Thiemann:1994pj, Thiemann:1992jj}). This is the idea we want to quickly suggest and explore here: to define a simpler model where similar techniques related to complex numbers could be explored. Indeed such implementation have been explored in reduced settings but never in a full environment with diffeomorphism invariance and true holonomy operators.

The idea can be stated as follows: in our model, we also have twice the amount of degrees of freedom from the standard theory. We could therefore regroup these degrees of freedom into complex variables and implement the constraints on the scalar product. In practice, there are several choices for the regrouping and only a full study of the possible discretization and the naturalness of the constraints in such a language could select one or the other. To illustrate the idea, we will just present one possibility, but others are available and could be useful.

A simple case would insist on the Poisson commutativity of the connection. We could also ask for $A$ and $X$ to be integrated along the lines of the support graph and $B$ and $e$ to be integrated along the dual lines. This leads to the following regrouping:
\begin{eqnarray}
A &=& \omega + \mathrm{i}X, \\
P &=& \frac{1}{2}\left(B - \mathrm{i}e\right).
\end{eqnarray}
$A$ indeed commutes with itself and it conjugate to $P$. We loose however the Poincaré connection we developed earlier. Modifications might be done to alleviate this problem, either by changing the variables or introducing a cosmological constant, maybe both. The corresponding reality constraints would read:
\begin{eqnarray}
\overline{A} &=& A, \\
\overline{P} &=& \mathrm{i} P.
\end{eqnarray}
These constraints would lead to constraints on the adjoint in the quantum theory and therefore on the scalar product.

The problem is simpler that the one in the full theory with regard to several points. First, there is no appearance of the spin connection compatible with the triad $\Gamma(E)$. All equations are here local which simplify things tremendously. The problem is also inherently $\mathrm{SU}(2)$ gauge-invariant and not $\mathrm{SL}(2,\mathbb{C})$ gauge-invariant. This might simplify the construction. And finally, the result is known, as we must reproduce the standard theory, which should guide research and help implementing a scalar product.

The problem is not trivial either and should help understand some points that we should find in the full theory. In particular, with $A$ real, it is easy to see that the holonomy operators when exponentiated will not preserve the condition on the momentum $P$. In the context of LQC, this has led to the definition of generalized holonomies \cite{Wilson-Ewing:2015lia} whose status is still rather unclear in the full theory especially regarding gauge-invariance. This setting should be sufficient to make this point more clear and explore how it can be resolved.

It should be noted however that some difficulties are lost. Most notably, the constraint on $A$ does not involve $P$ and the constraint on $P$ does not involve $A$. This complete separation of variables is a problem if we want to explore difficulties linked to the appearance of the spin connection in the full theory. In particular difficulties like the ordering problems are completely omitted. It could be possible however to restore some difficulty by implementing a cosmological constant. This cosmological constant will lead to the natural connection $\omega + \mathrm{i}e$ if negative for instance and might therefore involve some mixing capable of reproducing the problems we are interested in.

\section{Conclusion}

In this paper, we defined and developed a 3d toy-model for quantum gravity that reproduces some aspects of the simplicity constraints that appear in 3+1d. We have explored the model classically, discussed possible discretizations and defined a possible quantum version acting as a proof of concept showing that indeed, we can define natural second class constraints in 3d quantum gravity still leading to the same theory as the usual framework. The model could be developed for its own sake, for instance by studying variations on it, like introducing a cosmological constant, in order to better understand the new constraints and their role. This could be potentially interesting if the formalism was extended to higher dimensions. But our focus was really on having an equivalent of the reality conditions for self-dual variables in a simpler setting, compared to full Loop Quantum Gravity, but still complete enough to study techniques like the development of a continuum limit or the implementation of second class constraints in a fully diffeomorphism invariant theory. The goal of this paper therefore is to advocate for a reexamination of the self-dual variables or covariant variables in Loop Quantum Gravity by studying increasingly complicated models, starting with model as simple as Loop Quantum Cosmology, continuing with models more complicated like the one we are presenting and gradually including difficulties in order to reach the full theory.

\smallskip

The precise model we offered was developed by insisting on keeping the spin connection and the vielbein as independent variables and by defining their canonical momenta which we called $B$ and $X$. New constraints appear therefore to remove these extra momenta and translate the usual fact that the vielbein could directly be understood as the conjugated momenta to the connection. These constraints appear as second class constraints. This is a problem as second class constraints are notoriously difficult to implement in a diffeomorphism invariant quantum theory. This is also an opportunity as the model can act as a stepping stone towards more difficult constraints.

Classically and in the continuum, the model is straightforward to study. But, when discretizing the system, for purpose of loop quantization, several ambiguities arise. We unraveled some of them as it appeared that the constraints favored one discretization.  For the quantum version of the model, we proceeded in two steps. First we selected a method for quantizing the second class constraints. This was done by gauge-unfixing the set of constraints leading to a set of first class constraints that can be dealt with as usual in the loop quantization framework. This gauge unfixing was particularly easy in our case due to the simplicity of our 3d model. After the gauge unfixing, the definition of the Hilbert space and of the observables was rather straightforward. One subtlety arose with the choice of measure for the scalar product but this is due to the particularities of using a non semi-simple group. Though the quantization process was straightforward, it served as proof that the model such defined could be made sense of at the quantum level and be equivalent to standard 3d quantum gravity. Therefore further explorations, especially regarding possible implementation of reality conditions should be worthwhile and could lead to well-defined theories.

\smallskip

We want to conclude with this particular point: the goal of this model is not to explore new variables for quantum gravity but to provide a stepping stone for more difficult problems involving second class constraints or reality conditions. Our ultimate goal is to develop covariant Loop Quantum Gravity, either using self-dual variables or by having a new representation for a Lorentz covariant algebra of observables. The intermediate goal is to have a simpler model that should help develop techniques to implement reality conditions in diffeomorphism invariant quantum theories. The goal of this paper was to prove that such models exist and that it is reasonable to think that they will make sense in the quantum context.

\section*{Acknowledgements}

Many thanks to Etera Livine for the input he provided. Thank you also to Karim Noui for the very fruitful discussions which sparked the initial idea for this paper.

\appendix

\section{Constraint analysis for the 3+1d gravity model}
\label{app:constraints4d}

In this appendix, we explore the constraint analysis for the 3+1d model, with the tetrad and connection considered as independent variables. The analysis is similar to the one in \cite{Buffenoir:2004vx} which however considered the Plebanski theory. It is interesting to compare the results as we move along.

\subsection{Definitions}

\subsubsection{Sign convention}

In what follows, we work with the signature $(+\,-\,-\,-)$. For a given canonical variable (used in the Lagrangian) which we denote $x$ and its conjugated momentum $p$, we choose the following sign convention for the Poisson brackets: $\{p,x\} = 1$.

\subsubsection{Action}

We start with the usual 3+1D action based on the standard Lagrangian with an additional Holst term. Precisely, we have:
\begin{equation}
S[e,A] = -\frac{1}{16\pi G} \int_{\mathcal{M}} (\epsilon_{IJKL} + \frac{1}{\gamma} (\eta_{IK} \eta_{JL} - \eta_{IL} \eta_{JK})) e^I \wedge e^J \wedge F[A]^{KL}.
\end{equation}
$\mathcal{M}$ is a differential manifold isomorphic to $\mathbb{R}\times \Sigma$ where $\Sigma$ is a three-dimensional, compact, and borderless manifold. $\gamma$ is a free real parameter which we will call the Immirzi-Holst parameter. $e$ is the tetrad (vierbein) on the manifold $\mathcal{M}$ and $F$ is the curvature of $A$, a $\mathfrak{so}(3,1) \simeq \mathrm{sl}(2,\mathbb{C})$-valued connection. We can rewrite the previous action in a more detailed manner:
\begin{equation}
S[e,A] = -\frac{1}{8\pi G} \int_\mathbb{R} \mathrm{d}t \int_{\Sigma} \mathrm{d}^3 x (\epsilon_{IJKL} + \frac{1}{\gamma} (\eta_{IK} \eta_{JL} - \eta_{IL} \eta_{JK})) \epsilon^{abc} \left[ e_0^I e^J_a F[A]^{KL}_{bc} + e_a^I e_b^J F[A]^{KL}_{0c} \right] ,
\end{equation}
with:
\begin{equation}
F[A]^{IJ}_{\mu \nu} = \partial_\mu A_\nu^{IJ} - \partial_\nu A_\mu^{IJ} + (A_\mu^{IM} A_\nu^{NJ} - A_\nu^{IM} A_\mu^{NJ}) \eta_{MN}.
\end{equation}
The summation are done from $0$ to $4$ for capital Latin indices, from $1$ to $3$ for small Latin indices and from $0$ to $4$ for Greek indices.

\subsubsection{Momenta, primary constraints}

We start the Hamiltonian analysis by defining the conjugated momenta to $e$ and $A$ which we will denote, respectively, $X$ and $B$. Therefore, we have for;
\begin{equation}
S = \int_\mathbb{R} \mathrm{d}t L,
\end{equation}
the following definitions:
\begin{eqnarray}
X^\mu_I (x) \equiv \frac{\delta L}{\delta \partial_0 e_\mu^I(x)} ,\\
B^\mu_{IJ} (x) \equiv \frac{\delta L}{\delta \partial_0 A_\mu^{IJ}(x)} .
\end{eqnarray}
As the system is maximally singular (all derivatives appear linearly), we get the following primary constraints:
\begin{eqnarray}
\forall x,\mu,I,\, X^\mu_I (x) &=& 0,  \\
\forall x,I,J,\, B^0_{IJ}(x) &=& 0, \\
\forall x,a,I,J,\, B^a_{IJ}(x) &=& -\frac{1}{8\pi G} (\epsilon_{I'J'K'L'} + \frac{1}{\gamma} (\eta_{I'K'} \eta_{J'L'} - \eta_{I'L'} \eta_{J'K'})) \epsilon^{a'b'c'} e_{a'}^{I'}(x) e_{b'}^{J'}(x) \delta_{c'}^a (\delta^{K'}_I \delta^{L'}_J - \delta^{K'}_J \delta^{L'}_I).
\end{eqnarray}
To simplify this writing, we note:
\begin{equation}
\chi^{\gamma}_{IJKL} = \epsilon_{IJKL} + \frac{1}{\gamma} (\eta_{IK} \eta_{JL} - \eta_{IL} \eta_{JK}).
\end{equation}
We can then write the last constraint as:
\begin{equation}
\forall x,a,I,J,\, B^a_{IJ}(x) = -\frac{1}{4\pi G} \chi^{\gamma}_{KLIJ} \epsilon^{abc} e_{b}^{K}(x) e_{c}^{L}(x).
\end{equation}

\subsubsection{Starting Hamiltonian}

We can now construct the Hamiltonian as the Legendre transform of the Lagrangian. We get:
\begin{equation}
H = \int_{\Sigma} \mathrm{d}^3 x \left(\dot{e}_0^I (x) X^0_I(x) + \dot{e}_a^I(x) X^a_I(x) + \frac{1}{2} \dot{A}_0^{IJ}(x) B^0_{IJ}(x) + \frac{1}{2} \dot{A}_a^{IJ}(x) B^a_{IJ}(x) \right) - L
\end{equation}
where we used the more lightweight notation $\dot{x} \equiv \partial_0 x$. The $\frac{1}{2}$ factor compensate for the overcounting due to the antisymmetry of $A$ and $B$. Now, by replacing $L$ by its expression, we can find the following expression for $H$:
\begin{eqnarray}
H &=& \int_{\Sigma} \mathrm{d}^3 x \left(\dot{e}_0^I (x) X^0_I(x) + \dot{e}_a^I(x) X^a_I(x) + \frac{1}{2} \dot{A}_0^{IJ}(x) B^0_{IJ}(x)\right) \\
&+& \int_{\Sigma} \mathrm{d}^3 x \frac{1}{2} \dot{A}_a^{IJ}(x) \left(B^a_{IJ}(x) + \frac{1}{4\pi G} \chi^{\gamma}_{KLIJ} \epsilon^{abc} e_{b}^{K}(x) e_{c}^{L}(x) \right) \nonumber \\
&+& \frac{1}{8\pi G} \int_{\Sigma} \mathrm{d}^3 x \left( e_0^I(x) \chi^\gamma_{IJKL} \epsilon^{abc} e_a^J(x) F[A]^{KL}_{bc}(x) \right) \nonumber \\
&+& \frac{1}{8\pi G} \int_{\Sigma} \mathrm{d}^3 x \left( A_0^{IJ}(x) \epsilon^{abc}\left[ \partial_c (\chi^\gamma_{KLIJ} e_a^K(x) e_b^L(x) ) +(A_c^{NM}(x)\eta_{JN} \chi^\gamma_{KLIM} - A_c^{NM}(x)\eta_{MI} \chi^\gamma_{KLNJ}) e_a^K(x) e_b^L(x) \right] \right). \nonumber
\end{eqnarray}
The first two lines correspond to the primary constraints, with the derivative with respect to time as Lagrange multipliers. We can recognize on the third line the Einstein equations with a $e_0^I$ factor and the Gauß constraints on the forth line with a $A_0^{IJ}$ factor.

\subsection{Derivation of the constraints}

\subsubsection{Secondary constraints - equations of motion}

Preserving the primary constraints imply secondary constraints. We can quickly find that preserving $X^0_I(x)$ gives:
\begin{equation}
0 = D_I(x) \equiv \chi^\gamma_{IJKL} \epsilon^{abc} e_a^J(x) F[A]^{KL}_{bc}(x).
\end{equation}
Similarly, preserving $B_0^{IJ}(x)$ gives:
\begin{equation}
0 = G_{IJ}(x) \equiv \epsilon^{abc}\left[ \partial_c (\chi^\gamma_{KLIJ} e_a^K(x) e_b^L(x) ) +(A_c^{NM}(x)\eta_{JN} \chi^\gamma_{KLIM} - A_c^{NM}(x)\eta_{MI} \chi^\gamma_{KLNJ}) e_a^K(x) e_b^L(x) \right].
\end{equation}

\subsubsection{Secondary constraints - simplicity constraints}

The other secondary constraints are more difficult to handle. They come from the preservation of $X^a_I(x)$ and of the constraint on $B^a_{IJ}(x)$. Let's start by studying the conservation of $X^a_I(x) = 0$:
\begin{equation}
\{H, X^a_I(x)\} = \frac{1}{4\pi G} \epsilon_{abc} \chi^\gamma_{IJKL} \left[ \dot{A}^{KL}_b(x) e^J_c(x) + e^J_c(x) \nabla_c A^{KL}_0(x) + \frac{1}{2} e_0^J(x) F[A]_{bc}^{KL}(x) \right] .
\end{equation}
The appearance of the time derivative of $A$ suggests that the constraint will be of second class. Indeed, the tensor $\epsilon_{abc} \chi^\gamma_{IJKL} e^J_c(x)$ at a given $x$ seen as a rectangular matrix with columns labelled by $K$, $L$ and $b$ and lines labelled by $I$ and $a$ is of maximum rank (provided the tetrad is not degenerate). Therefore, it is always possible to find a collection of $\dot{A}^{KL}_b(x)$ such that the previous quantity vanishes. This of course does not fix the $\dot{A}$ factors completely but leaves six degrees of freedom that we expect to be fixed by further analysis.

This first constraint could be handled separately as the inversion problem completely decouples from the one for $B^a_{IJ}(x)$. We want now to compute the bracket between $H$ and $S^a_{IJ}(x) = B^a_{IJ}(x) + \frac{1}{4\pi G} \chi^{\gamma}_{KLIJ} \epsilon^{abc} e_{b}^{K}(x) e_{c}^{L}(x)$. To make things clearer, we'll do that in two steps. First:
\begin{eqnarray}
\{H, B^a_{IJ}(x)\} &=& \frac{1}{4\pi G} \left( \chi^\gamma_{MNKL} \epsilon^{dba} \nabla_b \left(e^M_0(x) e^N_d(x)\right) (\delta^K_I \delta^L_J - \delta^K_J \delta^L_I)  \right) \\
&-& \frac{1}{8\pi G} \left( A_0^{PQ}(x) \epsilon^{dba} e^K_d(x) e^L_b(x) \left[ \eta_{QN} \chi^\gamma_{KLPM} - \eta_{MP} \chi^\gamma_{KLNQ} \right] (\delta^N_I \delta^M_J - \delta^N_J \delta^M_I) \right). \nonumber
\end{eqnarray}
This can be simplified into:
\begin{eqnarray}
\{H, B^a_{IJ}(x)\} &=& \frac{1}{2\pi G} \left( \chi^\gamma_{MNIJ} \epsilon^{abc} \nabla_c \left(e^M_0(x) e^N_b(x)\right)  \right) \\
&-& \frac{1}{4\pi G} \left( A_0^{PQ}(x) \epsilon^{abc} e^K_b(x) e^L_c(x) \eta_{QN} \chi^\gamma_{KLPM} (\delta^N_I \delta^M_J - \delta^N_J \delta^M_I) \right). \nonumber
\end{eqnarray}
Now, we do the second computation:
\begin{equation}
\{H, \frac{1}{4\pi G} \chi^{\gamma}_{KLIJ} \epsilon^{abc} e_{b}^{K}(x) e_{c}^{L}(x)\} = \frac{1}{4\pi G} \chi^{\gamma}_{KLIJ} \epsilon^{abc} \left( \dot{e}_b^K(x) e^L_c(x) + e_b^K(x) \dot{e}^L_c(x) \right).
\end{equation}
As we mentioned earlier, the quantity $\dot{A}$ does not appear at all and the two conservations can be dealt with separately. Then, we have $3\times6 = 18$ constraints to be conserved but only $3\times4 = 12$ components for $\dot{e}$. We expect that $6$ new constraints should pop out. This is consistent with the fact that $6$ components of $\dot{A}$ were left unfixed.

Our goal is to find the $6$ combinations of the previous constraints that do not involve $\dot{e}$. We don't have to work with the full expression, we can focus on the following terms:
\begin{equation}
\Xi^a_{IJ}(x) \equiv \frac{1}{2} \chi^{\gamma}_{KLIJ} \epsilon^{abc} \left( \dot{e}_b^K(x) e^L_c(x) + e_b^K(x) \dot{e}^L_c(x) \right) = \chi^{\gamma}_{KLIJ} \epsilon^{abc} \dot{e}_b^K(x) e^L_c(x).
\end{equation}
We want to remove all the $\dot{e}$ appearances whatever their value. This means, we want to find the combinations of $\chi^{\gamma}_{KLIJ} \epsilon^{abc} e^L_c(x)$ which identically vanishes. Here is the solution:
\begin{equation}
\zeta^{k\ell b}_K(x) \equiv \chi^{-\frac{s}{\gamma}}_{MNPQ}e^M_p e^N_q \epsilon^{(kpq} \eta^{PI} \eta^{QJ} \chi^{\gamma}_{KLIJ} \epsilon^{\ell)bc} e^L_c(x) .
\end{equation}
$s$ is the signature of the space-time ($+1$ for Euclidean and $-1$ for Lorentzian). The quantity is symmetric in $k$ and $\ell$ so that for $b$ and $K$ fixed, there are $6$ different combinations. We can check that indeed $\zeta^{k\ell b}_K(x) = 0$. Let's start by computing:
\begin{eqnarray}
\chi^{-\frac{s}{\gamma}}_{MNPQ} \chi^\gamma_{KLIJ} \eta^{PI} \eta^{QJ} &=& \left(\epsilon_{MNPQ} - s \gamma\vphantom{\frac{1}{\gamma}} (\eta_{MP}\eta_{NQ} - \eta_{MQ}\eta_{NP}) \right) \left(\epsilon_{KLIJ} + \frac{1}{\gamma} (\eta_{KI}\eta_{LJ} - \eta_{KJ}\eta_{IL}) \right) \eta^{PI} \eta^{QJ} \\
&=& 2s(\eta_{MK} \eta_{NL} - \eta_{ML} \eta_{NK}) - 2s(\eta_{MK} \eta_{NL} - \eta_{ML} \eta_{NK}) + \frac{2}{\gamma} \epsilon_{MNKL} - 2s\gamma \epsilon_{KLMN} \nonumber \\
&=& 2\left(\frac{1}{\gamma} - s\gamma \right) \epsilon_{MNKL}. \nonumber
\end{eqnarray}
Let's not here that the special values of $\gamma = \pm 1$ in Euclidean space-times and $\gamma = \pm \mathrm{i}$ in Lorentzian space-times are singled out and correspond to non-invertibility cases. As we are interested in the Lorentzian case with real Immirzi, we will ignore such cases. We can turn back to the full expression. The previous quantity is contracted with three tetrads, we will therefore use the following property:
\begin{equation}
\epsilon_{MNKL} e^M_p e^N_q e_c^L \propto \epsilon_{pqc} \epsilon^{xyz} \epsilon_{MNKL} e^M_x e^N_y e_z^L ,
\end{equation}
which just comes from the fact that the space of $3$-forms in 3d is one dimensional. We then see the following contraction appearing:
\begin{equation}
\epsilon^{(kpq} \epsilon^{\ell)bc}\epsilon_{pqc} = 2\epsilon^{(k\ell)b} = 0.
\end{equation}
And therefore:
\begin{equation}
\zeta^{k\ell b}_K(x) = 0.
\end{equation}
This implies in turn that the following secondary constraints hold:
\begin{equation}
\Theta^{ab} = \chi^{-\frac{s}{\gamma}}_{MNPQ}e^M_p e^N_q \eta^{PI} \eta^{QJ} \epsilon^{(apq} \{H,B^{b)}_{IJ}\} = 0.
\end{equation}
We can compute this explicitly:
\begin{eqnarray}
\Theta^{ab} &=& \frac{1}{4\pi G}\chi^{-\frac{s}{\gamma}}_{MNPQ}e^M_p e^N_q \eta^{PI} \eta^{QJ} \epsilon^{(apq} \Big[ 2\left( \chi^\gamma_{KLIJ} \epsilon^{b)cd} \nabla_d \left(e^K_0(x) e^L_c(x)\right)  \right) \\
&-&  A_0^{RS}(x) \epsilon^{bcd} e^K_c(x) e^L_d(x) \eta_{SV} \chi^\gamma_{KLRU} (\delta^V_I \delta^U_J - \delta^V_J \delta^U_I) \Big] .  \nonumber
\end{eqnarray}
This simplifies a lot. It can be checked explicitly (though it is tedious) that the factor to $A_0^{RS}$ is symmetric and therefore the contraction vanishes. Similarly, applying the Leibniz rule and the same computation as above shows that the $e_0^K$ factor can come out of the derivative. We finally have:
\begin{equation}
\Theta^{ab} = \frac{1}{\pi G} \left(\frac{1}{\gamma} - s\gamma\right) \epsilon_{MNKL} e^M_p e^N_q  \epsilon^{(apq} \epsilon^{b)cd} e^K_0(x) \nabla_d e^L_c(x) .
\end{equation}
The interpretation of this is fairly straightforward: these are the remaining degrees of freedom of the torsion left out of the Gauß constraint as it is proportional to $e^{(a}_I T^{b) I}$, where $T$ is the torsion.

The presence of $e_0^I$ might be a bit unsettling but is not a big issue. Indeed, the previous still holds if we complete the three tetrads $e_a^I$ not with $e_0^I$ but with $e^0_I \propto \epsilon_{IJKL} e^J_a e^K_b e^L_c \epsilon{abc}$. Replacing $e_0$ with the previous construction in the constraint, we still get the remaining degrees of freedom of the torsion left out of the Gauß constraint but expressed as $g^{(ac} e^I_c T^{b) J} \eta_{IJ}$. It should be noted that it is trivial that the new constraint commutes with $X^0$ which incidentally proves that the previous one commutes with it two. We will continue with this new constraint to avoid carrying $e_0$ around.

\subsubsection{Tertiary constraints}
We don't expect any tertiary constraints. Indeed, a counting argument shows that we already have enough constraints to get $2$ degrees of freedom. And indeed, the conservation of the additional constraint $\Theta^{ab}$ gives only $6$ additional equations that fix the remaining degrees of freedom from $\dot{A}$, provided the tetrad is invertible. We have not explored the case of non-invertible tetrad, which is more complicated and we leave it to future investigation. Apart from this technical problem, the list of constraints from the Hamiltonian analysis is over. All that remains is the classification of constraints into first and second class.

\subsection{Classification of constraints}

None of the constraints expressed so far, except for the $X_I^0 = 0$ and $B_{IJ}^0 = 0$, are first class. As $A^{IJ}_0$ does not appear in any constraint, the $B_{IJ}^0$ case is trivial. As mentioned before, it is not trivial that $X^0_I$ commutes with $\Theta^ {ab}$ but the brackets only involves terms proportional to the torsion which vanishes if the tetrad is inverted. Anyway, as we are using our deformed constraint that does not involve $e_0$, we will rightly ignore this point. The other primary constraints are second class, as expected, but so are the Einstein equations of the quasi-Gauß constraint we found. We must therefore look for combinations of these to identify the actual first class constraints.

\subsubsection{Gauß constraints - first class}

Let's start with the simplest case: the Gauß constraints. It is the simplest case because thanks to our experience in 2+1d, we know what to expect. In effect, we expect these constraints should generate the local Lorentz transformation and this implies their form:
\begin{equation}
\mathcal{G}_{IJ} = \nabla_a B^a_{IJ}  + e_a^{[I} X^a_K \eta^{KJ]}
\end{equation}
This constraint is indeed a linear combination of $G_{IJ}$, the simplicity constraints, $X^0$ and $B^0$ and is, therefore, our correct Gauß constraint.

\subsubsection{Spatial diffeomorphism constraints - first class}

This case is more difficult than the three dimensional case. Indeed, as the tetrad appears in the equation of motions, they do not in general commute with the $X^a_I = 0$ constraints. It is a difficult problem in general but if we concentrate on the spatial diffeomorphism constraints and forget about the Hamiltonian constraint for now, we can once again be guided by the expected action of the constraints.

For the diffeomorphism constraints, we expect something of the following form:
\begin{equation}
\mathcal{D}_a = B^b_{IJ} F[A]^{IJ}_{ba} + \textrm{other terms not involving }B .
\end{equation}
This is because the first term correctly produces the diffeomorphism transformations for $A$ and $B$. On shell, the first term is proportional to:
\begin{equation}
B^b_{IJ} F[A]^{IJ}_{ba} \propto \chi^\gamma_{IJKL} e^K_c e^L_d \epsilon^{bcd} F^{IJ}_{ba} .
\end{equation}
This suggests that we should look for expressions quadratic in $e$ constructed out of $D_I$. The natural possibility is to consider:
\begin{equation}
\tilde{D}_a = e_a^I D_I .
\end{equation}
Subtracting a term proportional to $S^b_{IJ} F[A]^{IJ}_{ba}$ where $S$ is the simplicity constraint, we get the first term we were interested in. From there, they are two possibilities: either we continue constructing using the known transformation rules or we use the desired commutativity with other constraints to adjust the new terms. Either way, we get:
\begin{equation}
\mathcal{D}_a = B^b_{IJ} F[A]^{IJ}_{ba} + 2\pi G e_b^I \nabla_a X^b_I ,
\end{equation}
where $T_A[e]$ is the torsion of $e$ with respect to $A$.

\subsubsection{Hamiltonian constraint - first class}

Similarly, we can infer the correct Hamiltonian constraint from the expect result. In LQG with self-dual variables, the constraint normally reads:
\begin{equation}
\mathcal{H}_{LQG} \propto B^a_{IJ} B^b_{KL} F^{JL}_{ab} \eta^{IK} .
\end{equation}
We neglected prefactor that are not relevant to our discussion and depend on the density we choose for the Hamiltonian constraint. We want the simplest constraint possible and therefore take the density (2) that leads to a polynomial constraint. The form we have suggest that we consider the quantity:
\begin{equation}
\tilde{\mathcal{H}} = \chi^{\gamma}_{IJKL} \epsilon_{abc} e_a^J e_b^K e_c^L D_M \eta^{IM} .
\end{equation}
On-shell, thanks to the simplicity constraint, this reduces to the expression we were looking for. All we need is to adjust this expression so that it commutes with all the constraints, as we did for the spatial diffeomorphisms. The final result is:
\begin{equation}
\mathcal{H} = B^a_{IJ} B^b_{KL} F^{JL}_{ab} \eta^{IK} + 2\pi G \epsilon^{bcd} e_b^K e_c^L e_d^J \chi^\gamma_{MNKL} \chi^\gamma_{M'N'I'J} \eta^{II'} \eta^{MM'} \eta^{NN'} \nabla_a X^a_I .
\end{equation}

\subsubsection{Second class constraints}

The remaining constraints are all second class. Their form do not matter so much. So we will stick with the ones we have as there geometrical meaning is quite straightforward.

\subsection{Summary of the Hamiltonian analysis}

We have found the complete system of constraints, provided two hypotheses. We restricted to the case where the tetrad was invertible and to the case where the Immirzi-Holst parameter $\gamma$ did not correspond to the (anti-)self-dual case.

In that case, the system is defined by the following first class constraints:
\begin{equation}
\left\{\begin{array}{rcl}
X^0_I &=& 0 , \\
B^0_{IJ} &=& 0 , \\
\mathcal{G}_{IJ} &=& \nabla_a B^a_{IJ}  + e_a^{[I} X^a_K \eta^{KJ]} + e_0^{[I} X^0_K \eta^{KJ]} + A_0^{[IK} B^0_{KL} \eta^{LJ]} = 0 , \\
\mathcal{D}_a &=& B^b_{IJ} F[A]^{IJ}_{ba} + 2\pi G e_b^I \nabla_a X^b_I , \\
\mathcal{H} &=& B^a_{IJ} B^b_{KL} F^{JL}_{ab} \eta^{IK} + 2\pi G \epsilon^{bcd} e_b^K e_c^L e_d^J \chi^\gamma_{MNKL} \chi^\gamma_{M'N'I'J} \eta^{II'} \eta^{MM'} \eta^{NN'} \nabla_a X^a_I .
\end{array}\right.
\end{equation}
with the additional second class constraints:
\begin{equation}
\left\{\begin{array}{rcl}
X^a_I &=& 0, \\
S^a_{IJ} &=& B^a_{IJ}(x) + \frac{1}{4\pi G} \chi^{\gamma}_{KLIJ} \epsilon^{abc} e_{b}^{K}(x) e_{c}^{L}(x) = 0, \\
\Theta^{ab} &=& \frac{1}{\pi G} \left(\frac{1}{\gamma} - s\gamma\right) \epsilon_{MNKL} e^M_p e^N_q  \epsilon^{(apq} \epsilon^{b)cd} e^K_0(x) \nabla_d e^L_c(x) = 0 .
\end{array}\right.
\end{equation}
It should be noted that the Poisson brackets between the second class constraints are not ultra-local (they involve derivative of the Dirac delta). Techniques similar to one employed in \cite{Buffenoir:2004vx} should be considered if one wants to compute the Dirac brackets.

We left several problem for future investigation. One of the most salient point is the assumption of the invertibility of the tetrad. It is not clear that the current analysis survives non-invertibility for two reasons. First, the possibility of removing $e_0$ from the equations heavily relied on invertibility. Second, we know that the torsion might not vanish in certain case if the tetrad cannot be inverted \cite{Kaul:2016zbn}.

We also left the problem of attacking the quantum imposition of the constraints. This implies indeed some classical analysis whether we consider a gauge-unfixing strategy of a full Dirac procedure. We think that a Dirac procedure in particular might be a really good place to start as it is known that the tetrad formalism simplifies heavily some situations, for instance in bimetric theories of gravity \cite{Hinterbichler:2012cn, Chamseddine:2011mu}. This is however left open for further inquiries.

\bibliographystyle{bib-style}
\bibliography{biblio}

\end{document}